\documentclass[aps,prd,nopacs,nofootinbib,notitlepage]{revtex4-2}

\usepackage{amsfonts}
\usepackage{amsmath}
\usepackage{xcolor}
\usepackage{bm}
\usepackage{fancybox}
\usepackage{comment}
\usepackage{multirow}
\usepackage{tipa}
\usepackage{longtable}
\definecolor{refs}{RGB}{245,156,74}
\usepackage[colorlinks=true,hyperfootnotes=true,citecolor=cyan]{hyperref}

\usepackage{amsmath}
\usepackage{amssymb}
\usepackage{graphicx}
\usepackage{amsfonts}
\usepackage{latexsym}
\usepackage{bbold}
\usepackage{wasysym}
\usepackage{calligra}
\usepackage{float}
\usepackage{ulem}
\usepackage{inputenc}
\usepackage{xspace}
\usepackage{url}
\usepackage{epstopdf}
\usepackage{tikz}
\usepackage{amsthm}
\usepackage{soul}
\usepackage{hyperref}

\usepackage{amsmath}
\usepackage{graphicx}
\usepackage{amsfonts}
\usepackage{latexsym}
\usepackage{bbold}
\usepackage{wasysym}
\usepackage{calligra}
\usepackage{float}
\usepackage{ulem}
\usepackage{inputenc}
\usepackage{xspace}
\usepackage{url}
\usepackage{epstopdf}
\usepackage{tikz}

\usepackage{amsmath}
\usepackage{graphicx}
\usepackage{amsfonts}
\usepackage{latexsym}
\usepackage{bbold}
\usepackage{wasysym}
\usepackage{calligra}
\usepackage{float}
\usepackage{ulem}
\usepackage{inputenc}
\usepackage{xspace}
\usepackage{url}
\usepackage{epstopdf}
\usepackage{tikz}
\usepackage{amsthm}
\usepackage{enumitem}

\newcommand{\beq} {\begin{equation}}
\newcommand{\eeq} {\end{equation}}
\newcommand{\nb}{\nabla}
\newcommand{\sqg}{\sqrt{-g}}

\newcommand{\be}{\begin{equation}}
\newcommand{\ee}{\end{equation}}
\newcommand{\ba}{\begin{eqnarray}}
\newcommand{\ea}{\end{eqnarray}}
\newcommand{\bs}{\begin{subequations}}
\newcommand{\es}{\end{subequations}}

\newcommand{\diff}{\textrm{d}}

\newcommand{\lp}{\left(}
\newcommand{\rp}{\right)}
\newcommand{\lb}{\left[}
\newcommand{\rb}{\right]}

\begin{document}

\title{Hyperhydrodynamics: Relativistic Viscous Fluids from Hypermomentum}

	\author{Damianos Iosifidis}
    \email{damianos.iosifidis@ut.ee}
	\affiliation{Laboratory of Theoretical Physics, Institute of Physics, University of Tartu, W. Ostwaldi 1, 50411 Tartu, Estonia.}

\author{Tomi S. Koivisto}
\address{Laboratory of Theoretical Physics, Institute of Physics, University of Tartu, W. Ostwaldi 1, 50411 Tartu, Estonia}
\address{National Institute of Chemical Physics and Biophysics, R\"avala pst. 10, 10143 Tallinn, Estonia}
\email{tomi.koivisto@ut.ee}

\begin{abstract}

We explore a new action formulation of hyperfluids, fluids with intrinsic hypermomentum. Brown's Lagrangian for a relativistic perfect fluid is generalised by incorporating the degrees of freedom encoded in the hypermomentum tensor, namely by including connection-matter couplings. Quite interestingly,  generic hyperfluids are imperfect, since  hypermomentum induces such effects as bulk and shear viscosities as well as heat fluxes. The various coefficients that appear in the first order expansion of hydrodynamics can now be deduced from a Lagrangian formulation, given a geometrical interpretation and a suggested microscopic description in terms of hypermomentum.
This connection between hypermomentum and dissipative fluids could shed new light on the physics of relativistic hydrodynamics. 
The applicability of the new formalism is demonstrated by exact cosmological solutions.


\end{abstract}

\maketitle

\section{Introduction}

Gravity couples to the energy-momentum of matter. The possible couplings of matter to gravitational degrees of freedom beyond the metric are called hypermomentum\footnote{The hypermomentum tensor is formally defined as the variational derivative of the matter action wrt the independent affine connection \cite{hehl1976hypermomentumIII}.}. The well-known example is the coupling of spin-$\frac{1}{2}$ fermions to torsion in Poincar{\'e} models of gravity, see e.g. \cite{Hayashi:1980av,Trautman:1973wy,Magueijo:2012ug}. 
In the current gauge theory of gravity and spacetime, spin can couple to torsion universally\footnote{In the context of Poincar{\'e} models of gravity, ``the struggle raged for decades'', since the non-minimal coupling that had to be postulated for photons lead to an ``unphysical situation'' \cite{Hammond:2018man}.}, and thus all the matters known to the Standard Model of particle physics (barring the Higgs if assumed a scalar) are indeed ``hypermatters'' i.e. they can be  associated with nonzero hypermomentum \cite{Koivisto:2023epd}. 
This finding gives a fresh impetus to study hypermatters more generally. 

Hypermomentum has played a central role in the formulation of metric affine models of gravity \cite{hehl1995metric,hehl1976hypermomentum,hehl1978hypermomentum}. Such models comprise a framework that generalises Poincar{\'e} models of gravity by introducing non-metric degrees of freedom, which are associated to new, hypothetical microproperties of matter besides its spin, called dilaton and shear charges\footnote{There is an interesting possibility that the shear current might be related to the hadronic properties of matter \cite{hehl1997ahadronic}.}. This  geometrical framework remains of foundational importance in its original contexts in material physics such as continuum mechanics \cite{kroner}. Also, metric-affine geometry is a popular playing ground for modified gravity models, and recently there has been considerable progress in the field. The departure from general relativity can be achieved by either changing the gravitational Lagrangian by adding geometric invariants or by considering new connection-matter couplings. The former possibility, namely enlarging the gravitational sector of the action by adding more invariants to it may not be the safest road. For instance adding  quadratic curvature terms to the Einstein-Hilbert action 
can be problematic in general due to unphysical instabilities, see e.g.  \cite{Marzo:2021iok,Percacci:2020ddy,BeltranJimenez:2020sqf,Jimenez-Cano:2022sds} and special care must be given to the choice of the parameters in order to obtain healthy theories.
Instead of modifying the gravity action, reconsidering the coupling of the matter action in terms hypermomentum can be a more fruitful way of arriving at new gravitational dynamics\footnote{There is also an intriguing  conjecture regarding the quantum origin of hypermomentum \cite{Floerchinger:2021uyo}.}. For these reasons, it is interesting to take account also the possible dilaton and shear currents of hypermatters.


Dominating hypermomentum effects can occur in the very early Universe
\cite{Trautman:1973wy,Magueijo:2012ug} when the density of matter was taking on extremely large values, or in high density objects that may be scattered throughout the Universe, even compact objects \cite{Battista:2023znv,DeFalco:2023djo}. It is therefore quite natural to develop models of fluids that carry hypermomentum (hyperfluids) in a cosmological background. In \cite{Obukhov:1993pt,obukhov1996model} (see also \cite{Babourova:1995fv,babourova1998perfect}) one can find developments of hyperfluids which can be seen as generalizations of the Weyssenhoff spin fluid \cite{weyssenhoff1947relativistic}. However, as shown in \cite{Iosifidis:2020gth} the convective form of hypermomentum that these models allow for is not capable to fully reproduce the 5 degrees of freedom allowed by the isotropy demand \cite{minkevich1998isotropic,Boehmer:2006gd}. Therefore, the model developed in \cite{Iosifidis:2020gth} along with its generalization \cite{iosifidis2021perfect} are the appropriate ones to study isotropic and homogeneous cosmologies. (The Weyssenhoff fluid and its generalisations can allow nontrivial cosmological solutions when suitable approximations are justified \cite{Brechet:2008zz}.)

In order to have a better physical view and also guarantee the self-consistency of a fluid model one should be able to derive everything from an action principle. It is the purpose of this paper to formulate such a principle for the case of the perfect hyperfluid model \cite{Iosifidis:2020gth} by also dropping the homogeneity demand. 
The action formulation of relativistic perfect fluids by Brown uses variables which can be interpreted in terms of the physical properties of the fluid, and reduces to some previously considered formulations at the appropriate limits \cite{Brown:1992kc}. Previously, Brown's formulation has been extended with non-minimal couplings to scalar fields \cite{Boehmer:2015kta,Boehmer:2015sha,Koivisto:2015qua}
and to the metric \cite{Bettoni:2011fs,Bettoni:2015wla}.
Now the goal is to extend the action formulation for relativistic hyperfluids, which is tantamount to introducing connection-dependence into the fluid's action functional. It will turn out that the new couplings that generate hypermomentum alter also the energy-momentum tensor, giving it the form of an imperfect-viscous -fluid. What makes this interesting is that these properties of 
the fluid, such as its shear and bulk viscosity or heat fluxes, are determined by action principle and not implemented as ad hoc assumptions. It is thus possible to understand the origin of the viscous properties of the fluid in the microstructure of the medium, hypermomentum.


Relativistic hydrodynamics started with the seminal work of Eckart \cite{eckart1940thermodynamics}\footnote{A closely related model was also developed a few years later by Landau and Lifshitz in their fluid mechanics book \cite{landau2013fluid}}. However, Eckart's theory has some undesirable features, 1) it is not causal, 2) it is unstable and 3) for rotating fluids the initial value problem is not well-posed \cite{hiscock1985generic}. A generalization was developed by Israel \cite{israel1976nonstationary} which overcomes all these unattractive features. Hiscock and Lindblom later generalized Israel's theory by including non-uniform equilibrium configurations \cite{hiscock1983stability}.  The state of art in relativistic hydrodynamics is nowadays to understand it it as an effective field theory. More specifically, hydrodynamics is developed as a derivative expansion in temperature $T$, d-velocity $u^{\mu}$, and chemical potential $\mu$ that describe the fluid \cite{baier2008relativistic} (see also \cite{romatschke2019relativistic,Kovtun:2012rj,Romatschke:2009im,pandya2021numerical,bemfica2022first,rezzolla2013relativistic} for details). To zeroth order one has the perfect fluid and the gradients of $T, u^{\mu}$ and $\mu$ describe out of equilibrium configurations. Truncating the derivative expansion to first order\footnote{There is a region in the parameter space of generic first-order relativistic hydrodynamics for which  perturbations around the equilibrium are  stable \cite{kovtun2019first}.} one then describes viscous fluids, obtaining the Navier-Stokes equations. Keeping up to second derivative terms we have $2^{nd}$ order hydrodynamics and ascending up the ladder on the derivative expansion one is lead to higher-order dissipative hydrodynamics\footnote{There is an increasing interest in the study of the Carrollian limit of relativistic fluids (see for example \cite{Ciambelli:2018xat,Petkou:2022bmz}) }. 

Let us note that even though the derivative expansion approach is formal in nature, the coefficients functions of such an expansion are not evaluated directly (namely, they remain free) and their values must be obtained by a matching to some underlying microscopic theory. 
As we shall show in what follows, quite remarkably, in our development here the coefficient functions for the viscous terms are found exactly and are given in terms of the hypermomentum variables (i.e. microstructure). In a sense, in an attempt to excite hypermomentum, not only have we arrived to a dissipative fluid configuration but we are also able to  relate exactly the various coefficients of this expansion to the microscopic properties of the fluid as they described by the hypermomentum tensor. This suggests the inclusion of hypermomentum in the development of relativistic dissipative fluids. 

The paper is organized as follows. Firstly in sec. \ref{setup}, we communicate the basic ingredients (definitions, conventions etc) that are going to be used throughout. Then, in sec. \ref{dilation} we consider the perhaps simplest example, a dilation hyperfluid.    
In sec. \ref{spin} we investigate the spin hyperfluids in the new formalismn, and in sec. \ref{shear} we look at the generic couplings and focus in particular the remaining shear hypermomentum.
We then review the case of general hyperfluid, summarising the results of the previous derivations in sec. \ref{general}. The case of perfect hyperhydrodynamics is discussed in sec. \ref{perfect}. To demonstrate the applicability of the formalism, we derive examples of exact cosmological solutions sourced by different types of hyperfluids in sec. \ref{cosmo}. Finally, sec. \ref{conclu} presents a brief conclusion. Some technical details are given in the appendices. 

\section{The Setup}
\label{setup}

We shall work in the Metric-Affine framework. To this end we consider a d-dimensional\footnote{In applications, sec. \ref{cosmo}, we set $d=4$.} manifold of Lorentzian signature which we endow with a metric tensor $g=g_{\mu\nu}dx^{\mu}dx^{\nu}$, $\mu,\nu=0,1,2,...,d-1$ and a generic affine connection $\nabla$ with components $\Gamma^{\lambda}{}_{\mu\nu}$. We define the torsion and curvature of the connection according to
\beq
S_{\mu\nu}{}{}^{\lambda}:=\Gamma^{\lambda}{}{}_{[\mu\nu]}
\eeq
and 
\begin{equation}
R^{\mu}{}_{\nu\alpha\beta}:= 2\partial_{[\alpha}\Gamma^{\mu}{}_{|\nu|\beta]}+2\Gamma^{\mu}{}_{\rho[\alpha}\Gamma^{\rho}{}_{|\nu|\beta]}
\end{equation}
respectively. Obviously these objects need only a connection do be defined and do not require a metric. One the other hand a metric (along with the connection) is essential to construct the non-metricity tensor which we define via,
\beq
Q_{\alpha\mu\nu}=-\nabla_{\alpha}g_{\mu\nu}
\eeq
and measures the failure of the metric to be covariantly constant. Out of non-metricity we can construct 2 vectors by the following contractions
\beq
Q_{\mu}:=Q_{\mu\alpha\beta}g^{\alpha\beta}\quad, \quad q_{\mu}:=Q_{\alpha\nu\mu}g^{\alpha\nu}
\eeq
As for torsion we can define a vector along with a pseudo-vector, as follows
\beq
S_{\mu}:=S_{\mu\alpha}{}{}^{\alpha}\quad, \quad t_{\mu}:=\epsilon_{\mu\nu\alpha\beta}S^{\nu\alpha\beta}
\eeq
  with the latter being defined only for $n=4$. Note that with the above definitions of torsion and non-metricity the affine connection can be decomposed as
 (see for instance \cite{iosifidis2019metric}) 
\begin{equation}\label{decgamma}
{\Gamma^\lambda}_{\mu \nu} = \tilde{\Gamma}^\lambda_{\phantom{\lambda} \mu \nu} + {N^\lambda}_{\mu \nu}\,,
\end{equation}
where
\beq\label{lcconn}
\tilde{\Gamma}^\lambda_{\phantom{\lambda}\mu\nu} = \frac12 g^{\rho\lambda}\left(\partial_\mu 
g_{\nu\rho} + \partial_\nu g_{\rho\mu} - \partial_\rho g_{\mu\nu}\right)
\eeq
is the usual Levi-Civita connection and the additional tensor ${N^\lambda}_{\mu\nu}$ defined through
\beq\label{distortion}
{N^\lambda}_{\mu\nu} = {\frac12 g^{\rho\lambda}\left(Q_{\mu\nu\rho} + Q_{\nu\rho\mu}
- Q_{\rho\mu\nu}\right)} - {g^{\rho\lambda}\left(S_{\rho\mu\nu} +
S_{\rho\nu\mu} - S_{\mu\nu\rho}\right)} \,.
\eeq	
 is the so-called distortion tensor, measuring the departure from the Riemannian geometry.
 Out of the distortion we can extract three traces as $N^{(1)}_{\nu}:=N^{\alpha}{}_{\alpha\nu}$, $N^{(2)}_{\nu}:=N^{\alpha}{}_{\nu\alpha}$ and $N^{(3)}_{\nu}:=N_{\alpha\mu\nu}g^{\mu\nu}$ which are linear combinations of the torsion and non-metricity vectors as can be trivially verified.
 
 Using the above split (which is known as post-Riemannian expansion \cite{hehl1995metric}) of the affine connection, equation (\ref{decgamma}), we can decompose any quantity into the usual Riemannian part plus contributions associated to torsion and non-metricity. This proves to be very helpful in many instances and will be used in the current study in several places. In all such expansions quantities with a  $\widetilde{}$ will denote Riemannian parts namely the ones constructed only out of the Levi-Civita part of the connection.  Finally, let us note
that from the curvature tensor and without any metric  we can construct   two independent contractions as
\beq
R_{\nu\beta}:=R^{\mu}{}_{\nu\mu\beta}\quad,\quad \widehat{R}_{\alpha\beta}:=R^{\mu}{}_{\mu\alpha\beta}
\eeq
The former defines as usual the Ricci tensor while the latter is the so-called  homothetic curvature tensor and is of purely non-Riemannian origin. Once a metric is given we can form yet another  contraction
\beq
\breve{R}^{\lambda}{}_{\kappa}:=R^{\lambda}{}_{\mu\nu\kappa}g^{\mu\nu}
\eeq
which is oftentimes referred to as the co-Ricci tensor. It is important to stress that the Ricci scalar is still uniquely defined since
\beq
R:=R_{\mu\nu}g^{\mu\nu}=-\breve{R}_{\mu\nu}g^{\mu\nu}\quad,\quad \widehat{R}_{\mu\nu}g^{\mu\nu}=0
\eeq
Having  developed all the  necessary geometric setup let us now focus on the physics.

\subsection{Metric-Affine Conservation Laws}

The sources in Metric-Affine Gravity are the  Canonical and Metrical Energy-Momentum Tensors, that are respectively given by\footnote{Here $e_{\mu}{}^{A}$ is as usual the vielbein which defines the metric through $g_{\mu\nu}=e_{\mu}{}^{A}e_{\nu}{}^{B}g_{AB}$ where $g_{AB}$ is the tangent space metric. For orthonormal vielbeins, the latter becomes the flat Minkowski metric, i.e. $\eta_{AB}=(-1,1,1,...,1)$.}
\beq\label{cemt}
{t^\mu}_A := \frac{1}{\sqrt{-g}} \frac{\delta S_{\text{M}}}{\delta {e_\mu}^A} \,.
\eeq
and
\beq
T_{\mu \nu} := - \frac{2}{\sqrt{-g}} \frac{\delta S_{\text{M}}}{\delta g^{\mu \nu}} = - \frac{2}{\sqrt{-g}} \frac{\delta (\sqrt{-g} \mathcal{L}_{\text{M}})}{\delta g^{\mu \nu}} \,,
\eeq
 along with the Hypermomentum tensor  \cite{hehl1976hypermomentum,hehl1978hypermomentum}
 which is defined by as the variation of the matter part of the action with respect to the connection, viz.
\beq
{\Delta_\lambda}^{\mu \nu} := - \frac{2}{\sqrt{-g}} \frac{\delta S_{\text{M}}}{\delta {\Gamma^\lambda}_{\mu \nu}} = - \frac{2}{\sqrt{-g}} \frac{\delta (\sqrt{-g} \mathcal{L}_{\text{M}})}{\delta {\Gamma^\lambda}_{\mu \nu}} \,.
\eeq
The latter tensor is of ultimate importance since it encodes the micro-properties of matter such as spin, dilation and shear. It can be split into these its irreducible pieces of spin, dilation and shear according to (see \cite{hehl1995metric})
\beq
\Delta_{\alpha\mu\nu}=\Sigma_{\alpha\mu\nu}+\frac{1}{n}g_{\alpha\mu}\Delta_{\nu}+\hat{\Delta}_{\alpha\mu\nu} \label{hypsplit}
\eeq
with
\beq
\sigma^{\mu\nu\alpha}:=\Delta^{[\mu\nu]\alpha} \qquad (Spin)
\eeq
\beq
\Delta^{\nu}:=\Delta^{\alpha\mu\nu}g_{\alpha\mu} \qquad (Dilation)
\eeq
\beq
\Sigma^{\mu\nu\alpha}:=\Delta^{(\mu\nu)\alpha}-\frac{1}{n}g^{\mu\nu}\Delta^{\alpha} \qquad (Shear)
\eeq
In MAG framework, the matter action is assumed to possess the GL and Diffeomorphism invariances. These invariances give the conservation laws \cite{hehl1995metric,Iosifidis:2020gth} (expressed in a holonomic frame here):
	\beq
	t^{\mu}{}_{\lambda}
	= T^{\mu}{}_{\lambda}-\frac{1}{2 \sqrt{-g}}(2S_{\nu}-\nabla_{\nu})(\sqrt{-g}\Delta_{\lambda}{}^{\mu\nu}) \label{cc1}
	\eeq
	\beq
	\frac{1}{\sqrt{-g}}(2S_{\mu}-\nabla_{\mu})(\sqrt{-g}t^{\mu}{}_{\alpha})=-\frac{1}{2} \Delta^{\lambda\mu\nu}R_{\lambda\mu\nu\alpha}+\frac{1}{2}Q_{\alpha\mu\nu}T^{\mu\nu}+2 S_{\alpha\mu\nu}t^{\mu\nu} \label{cc2}
	\eeq
	This is the most general form. An alternative form may be obtained if we expand out the covariant derivative in terms of the Levi-Civita plus distortion. After some trivial algebra we find:
\beq
    2(t^{\mu}{}_{\lambda}-T^{\mu}{}_{\lambda})=\widetilde{\nabla}_{\nu}(\Delta_{\lambda}{}^{\mu\nu})+N^{\mu}{}_{\alpha\beta}\Delta_{\lambda}{}^{\alpha\beta}-N^{\alpha}{}_{\lambda\beta}\Delta_{\alpha}{}^{\mu\beta} \label{t}
\eeq
\beq
\widetilde{\nabla}_{\mu}t^{\mu}{}_{\alpha}=\frac{1}{2} \Delta^{\lambda\mu\nu}R_{\lambda\mu\nu\alpha}+\frac{1}{2}Q_{\alpha\mu\nu}(t^{\mu\nu}-T^{\mu\nu})+(Q_{\mu\nu\alpha}+S_{\mu\nu\alpha}-2 S_{\alpha\mu\nu})t^{[\mu\nu]}
\eeq
 The advantage of these last two expressions is that they both contain the Levi-Civita covariant derivative, meaning that we can freely move the metric in and out of them and perform the various contractions and raising/lowering of the indices. We can also fully eliminate $T^{\mu\nu}$ from (\ref{cc2}) by employing (\ref{cc1}), resulting in
\begin{gather}
\widetilde{\nabla}_{\mu}t^{\mu}{}_{\alpha}=\frac{1}{2} \Delta^{\lambda\mu\nu}R_{\lambda\mu\nu\alpha}+(Q_{\mu\nu\alpha}+S_{\mu\nu\alpha}-2 S_{\alpha\mu\nu})t^{[\mu\nu]}\nonumber \\
+\frac{1}{4}Q_{\alpha\mu\nu}\Big( \widetilde{\nabla}_{\lambda}\Delta^{\nu\mu\lambda}+N^{\mu\beta\gamma}\Delta^{\nu}{}_{\beta\gamma}-N^{\lambda\nu}{}{}_{\beta}\Delta_{\lambda}{}^{\mu\beta}\Big)
\end{gather}
The above conservation laws describe the evolution of the MAG sources\footnote{These conservation laws can be used to compute how ad hoc hypermatter moves in non-Riemannian backgrounds \cite{iosifidis2023motion}. For the coupling of metric-affine spacetime geometry to all matters known to the Standard Model, see \cite{BeltranJimenez:2020sih}.}. Alternatively, one may eliminate the canonical tensor using (\ref{cc1}) and arrive at
(see for instance \cite{Iosifidis:2020gth})
\begin{equation}\label{conslawenmomhyperm}
\sqrt{-g} \left( 2 \tilde{\nabla}_\mu {T^\mu}_\alpha - \Delta^{\lambda \mu \nu} R_{\lambda \mu \nu \alpha} \right) + \hat{\nabla}_\mu \hat{\nabla}_\nu \left( \sqrt{-g} {\Delta_\alpha}^{\mu \nu} \right) + 2 {S_{\mu \alpha}}^\lambda \hat{\nabla}_\nu \left( \sqrt{-g} {\Delta_\lambda}^{\mu \nu} \right) = 0 \,, 
\end{equation}
with
\begin{equation}\label{hatnabla}
\hat{\nabla}_\mu := 2 S_\mu - \nabla_\mu \,,
\end{equation}
which is the modified energy-momentum tensor conservation for microstructured matter.

\subsection{The Perfect Hyperfluid}

In General Relativity and also in the broad context of metric theories of gravity one is acquainted with the notion of a perfect fluid. The energy-momentum tensor of the perfect fluid has the usual isotropic form. Now that we also have hypermomentum, it is natural to ask what would then be the generalization of the perfect fluid with an isotropic hypermomentum included (i.e. a Hyperfluid). Such a fluid model was developed by one of us in \cite{Iosifidis:2020gth} (see also \cite{iosifidis2021perfect} for a generalization) So,
demanding  spatial isotropy  the hypermomentum tensor of the Perfect Hyperfluid takes the covariant form \cite{Iosifidis:2020gth}	
	\beq
	\Delta_{\alpha\mu\nu}=\phi h_{\mu\alpha}u_{\nu}+\chi h_{\nu\alpha}u_{\mu}+\psi u_{\alpha}h_{\mu\nu}+\omega u_{\alpha}u_{\mu} u_{\nu}+\delta_{d,4}\epsilon_{\alpha\mu\nu\kappa}u^{\kappa}\zeta \label{Dform}
	\eeq
where $\phi,\chi,\psi,\omega$ and $\zeta$ are the functions encoding the microscopic characteristics of the fluid\footnote{Note that the $\epsilon_{\alpha\mu\nu\kappa}u^{\kappa}\zeta$ part is non-vanishing only for $d=4$. For brevity in the following we shall omit the Kronecker $\delta_{d,4}$ factor.}.  Its 3 pieces of dilation, spin and shear read
\beq
\Delta_{[\alpha\mu]\nu}=(\psi-\chi)u_{[\alpha}h_{\mu]\nu}+\epsilon_{\alpha\mu\nu\kappa}u^{\kappa}\zeta \label{spin}
\eeq
\beq
\Delta_{\nu}:=\frac{1}{d}\Delta_{\alpha\mu\nu}g^{\alpha\mu}=\frac{1}{d}\Big[ (d-1) \phi -\omega\Big] u_{\nu} \label{dil}
\eeq
\beq
\breve{\Delta}_{\alpha\mu\nu}=\Delta_{(\alpha\mu)\nu}-g_{\alpha\mu}\Delta_{\nu} =\frac{(\phi+\omega)}{d}\Big[ h_{\alpha\mu}+(d-1)u_{\alpha}u_{\mu} \Big] u_{\nu} +(\psi +\chi)u_{(\mu}h_{\alpha)\nu} \label{shear}
\eeq
respectively. The physical meaning of $(\phi, \chi,\psi,\omega,\zeta)$ becomes more clear if we introduce the field redefinitions
\begin{equation}
h :=\frac{(d-1)}{d} \phi-\frac{1}{d}\omega \quad (dilation)
	\end{equation}
\beq
	\sigma:=(\psi-\chi),\qquad\zeta \quad (spin)
	\eeq 
	\beq
\Sigma:=	\frac{(\psi+\chi)}{2} , \qquad \Pi:=\frac{(\phi+\omega)}{d} \quad (shear)
	\eeq
	In the above, the first combination is the dilation,  in the second line we have the two spin parts and $\Sigma, \Pi$ correspond to the two parts of shear. With these redefinitions we may re-express the hypermomentum parts as
		\beq
	\Delta_{[\alpha\mu]\nu}= \sigma u_{[\alpha}h_{\mu]\nu}+\epsilon_{\alpha\mu\nu\kappa}u^{\kappa}\zeta \label{spinD}
	\eeq
	\beq
	\Delta_{\nu}:=\frac{1}{d}\Delta_{\alpha\mu\nu}g^{\alpha\mu}=hu_{\nu}   \label{dil}
	\eeq
	\beq
	\breve{\Delta}_{\alpha\mu\nu}=\Pi\Big[ h_{\alpha\mu}+(d-1)u_{\alpha}u_{\mu} \Big] u_{\nu} +2 \Sigma u_{(\mu}h_{\alpha)\nu} \label{shearDD}
	\eeq
For the last shear part it is also convenient to split it further as
\beq
\breve{\Delta}_{\alpha\mu\nu}=(\Sigma+\Pi)u_{\alpha}u_{\mu}u_{\nu}+\frac{1}{d}g_{\alpha\mu}\Sigma u_{\nu}+\frac{1}{2}g_{\nu\alpha}\Pi u_{\mu}+\frac{1}{2}g_{\nu\mu}\Pi u_{\alpha}
\eeq

Furthermore, introducing the  densitized hypermomentum fluxes 
\beq
\sigma^{\mu}=\sqrt{-g}\sigma u^{\mu} \quad , \quad \zeta^{\mu}:=\sqrt{-g}\zeta u^{\mu} \label{ssp}
\eeq
\beq
\mathcal{D}^{\mu}:=\sqrt{-g}h u^{\mu}
\eeq
\beq
\Sigma^{\mu}:=\sqrt{-g}\Sigma u^{\mu} \quad, \quad \Pi^{\mu}:=\sqrt{-g}\Pi u^{\mu} \label{ssp2}
\eeq
The above  hypermomentum parts are expressed as
\beq
\sqrt{-g}\Delta_{[\alpha\mu]\nu}= \sigma_{[\alpha}g_{\mu]\nu}+\epsilon_{\alpha\mu\nu\kappa}\zeta^{\kappa}
\eeq
\beq
\sqrt{-g}\Delta_{\nu}=\mathcal{D}_{\nu}
\eeq
\beq
\sqrt{-g}\breve{\Delta}_{\alpha\mu\nu}=(\Sigma_{\nu}+\Pi_{\nu})u_{\alpha}u_{\mu}+\frac{1}{d}g_{\alpha\mu}\Sigma_\nu+\Pi_{(\mu}g_{\alpha)\nu}
\eeq
which turn out to be quite convenient when performing variations\footnote{This is so because the densitized fluxes are by construction metric independent (see also discussion and proofs in the Appendix)}. The full hypermomentum density then reads
\begin{gather}
    \sqrt{-g}\Delta_{\alpha\mu\nu}= \sigma_{[\alpha}g_{\mu]\nu}+\epsilon_{\alpha\mu\nu\kappa}\zeta^{\kappa}+g_{\alpha\mu}\mathcal{D}_{\nu}+(\Sigma_{\nu}+\Pi_{\nu})u_{\alpha}u_{\mu}+\frac{1}{d}g_{\alpha\mu}\Sigma_\nu+\Pi_{(\mu}g_{\alpha)\nu} \label{FullHyper}
\end{gather}
We have now all the ingredients that we need in order to develop the Lagrangian formulation of the Hyperfluid. We shall start with the simplest case of a purely dilational Hyperfluid and gradually build it up to the generic full  Hyperfluid with all three parts excited.

We should remark, as it is probably clear from (\ref{ssp})-(\ref{ssp2}) that the 5 hypermomentum variables we are using are in fact hypermomenta per particle number n since for all 5 densities $\mathcal{X}^{\mu}=\{D^{\mu},\sigma^{\mu},\zeta^{\mu},\Sigma^{\mu},\Pi^{\mu}\}$ it holds that
\beq
\frac{\mathcal{X}^{\mu}}{|\mathcal{X}|}=\frac{J^{\mu}}{|J|}
\eeq
where $J^{\mu}=\sqrt{-g}n u^{\mu}$ is the densitized particle number flux vector and $|J|=\sqrt{-g_{\mu\nu}J^{\mu}J^{\nu}}$. Accordingly, the particle number density is given by
\beq
n:=\frac{|J|}{\sqrt{-g}}
\eeq
and similarly we have 
\beq
\mathcal{X}:=\frac{|\mathcal{X}|}{\sqrt{-g}}
\eeq
for the rest of hypermomentum variables. More information and technical tools related to the above exposure are given in Appendix A, and below we discuss the physical interpretation of the new hyperfluid model and compare it with the previous hyperfluid models in the literature.  

\subsection{On the new hyperhydrodynamic approach}

The fundamental principles of hydrodynamics, the laws of conservation of mass, momentum, and energy, are
extended in hyperhydrodynamics by the conservation of the 5 quantities characterising the hypermomenta. As clarified above, though we can associate currents to each of these hypermomentum degrees of freedom, the fundamental variational degrees of freedom are not the currents but the hypermomentum charges. As an example (to be considered in more detail below in section \ref{dilation}), we can extend the standard fluid description to incorporate also a dilational property of the fluid by endowing it with the dilation charge per particle, $h$. This implies that there is a dilaton potential of a fluid element, locally given by the ''dilation density'' $nh$, where $n$ is the number density. The associated dilation flux is $\mathcal{D}^{\mu}=\sqrt{-g}\Delta^{\mu}=\sqrt{-g}hn u^{\mu}$. It is a matter of choice whether to consider the dilation density $hn$ or alternatively the dilation per particle $h$ as the independent variable, since these prescriptions would lead to equivalent results, but we emphasise that it would not be consistent to consider the dilation flux $\mathcal{D}^{\mu}$ as a the fundamental variational degree of freedom, since it is not a scalar field but has 4 components. Physically, there cannot be dilation flux independently of the flux of particles, so we must have $\mathcal{D}^{\mu} \sim J^\mu$, but the implied proportionality can vary in space and time.  

Therefore, in this set-up, all the hypermomentous interactions are constructed from nonminimal couplings of the affine connection to the particle flux $J^\mu$, and they should be proportional to the hypermomentum densities such as $h$. Thus, in contrast to the previous formulations of hyperfluids, the hyperhydrodynamical approach incorporates hypermomentum at the effective level of hydrodynamics. In the traditional approach, following Cosserats, one considers the dynamics of a material frame, rigid for spin fluids and deformable for a hyperfluid with shear. Then, in addition to the fluid's four-velocity $u^\mu$, one has to introduce also the Bartels frame $\alpha_A{}^\mu$, i.e. the spatial components of the material frame whose time component can be considered the four-velocity $\alpha_0{}^\mu=u^\mu$, thus 
introducing yet 12 new variational degrees of freedom, which are further multiplied for technical reasons\footnote{Brown refers to the $(d-1)$ Lagrangian coordinates of the fluid $\alpha^A$. The Obukhov \& Tresguerres hyperfluid action \cite{Obukhov:1993pt} was instead exploiting the fluid coframe $\beta^A{}_\mu$, which is supposed to have an inverse
$\alpha_A{}^\mu$ such that $\beta^A{}_\mu\alpha_B{}^\mu=\delta^A_B$, see alternative variables in \cite{Obukhov:2023yti}. We will also drag along the $\alpha^A$ and the $\beta_A$ in our equations so that the model reduces to Brown's Lagrangian when hypermomentum is switched off, even though these variables play no essential role in our derivations.} (one considers the inverse material frame components as independent degrees of freedom, adds Lagrange multipliers to impose the inverse relation plus further Lagrange multipliers to impose the other relations required for consistency, see \cite{Obukhov:1993pt,obukhov1996model,Babourova:1995fv,babourova1998perfect} for more details). Essentially, besides the usual tetrad in gravity theory, another, material tetrad is invoked in the conventional hyperfluid gravity theory. Then one has at hand the fields of tetrad bigravity recently considered by Blixt {\it et al} \cite{Blixt:2023qbg}. This opens many interesting possibilities and allows to speculate on the possible connection of the material tetrad components to microscopic degrees of freedom, hypothetically emerging from a more fundamental description of matter fields ("manifields"). However, a simpler formulation is available for the consistent model of gravity sourced by a hyperfluid, based on the rather straightforward extension of the standard concepts in relativistic fluid dynamics. Since neither the conventional Cosseratian approach nor the new hyperhydrodynamical approach can entertain pretensions beyond an effective theory (meaning a macroscropic, averaged description of some elementary constituents of matter), the new and more minimal approach could turn out to be the more advantageous for some purposes. Whereas the conventional hyperfluid models have yet found very few concrete applications, we shall show in this article that the hyperhydrodynamical fluid is amenable to even exact solutions.   

Besides technical simplicity, there can also be physical motivations to consider the new hyperhydrodynamical alternative. As pointed out in the introduction, the previously studied hyperfluids, which can be seen as generalizations of the Weyssenhoff spin fluid \cite{weyssenhoff1947relativistic}, are well known to be incompatible with cosmological symmetry because they are subject to the so called Frenkel condition \cite{minkevich1998isotropic,Boehmer:2006gd,Brechet:2008zz,Iosifidis:2020gth}. 
However, the physical justification for the Frenkel condition is questionable. At a more fundamental level, since spin is the charge of the Lorentz group, matter with spin should be coupled to the proper Lorentz gauge theory of gravity. It then turns out that spinning matter does not respect the Frenkel condition\footnote{The non-vanishing components of the Lorentz gauge field compatible with cosmological symmetry are $\omega^{0i}{}_\mu d x^\mu = A dx^i$, $\omega^i{}_{j\mu} dx^\mu = iB\epsilon^i{}_{jk} dx^k$. If we interpret the $B$ as the rotation, the $A$ as the boost degree of freedom in geometry, spinning matter in rest frame is expected to excite the $B$.} \cite{Koivisto:2023epd}. Further, an important distinction hyperhydrodynamics approach wrt to the Cosseratian hyperfluid formulations is that the latter describe solely perfect fluids. In hydrodynamics, dissipative effects such as heat flows can arise from various phenomena. In hyperhydrodynamics, the fluid's flow is coupled to the affine spacetime geometry in a way which can exert an effective friction, such that the fluid's flow through a flow resistance causes dissipation. We indeed find that the hyperhydrodynamical couplings can induce effective heat flows and viscosity to the fluid. Thus, our generic hyperfluid is imperfect (we will discuss the perfect fluid sector in Section \ref{perfect}). This allows the description of a wider range of potential physical phenomena which are the plausible consequences of nonminimal couplings of matter fields and affine spacetime geometry. Ultimately, it has to be decided by experiments which hyperfluid model, if any, provides the realistic description of a given physical system. 

\section{Action Functional for Pure Dilation Hyperfluid}
\label{dilation}
Let us recall that the hypermomentum tensor of a pure dilation fluid has the form
\beq
\Delta_{\lambda}{}^{\mu\nu}=\delta_{\lambda}^{\mu}\Delta^{\nu}\,. \label{dil}
\eeq
It is not difficult to show that the variation of the term $Q_{\mu}\mathcal{D}^{\mu}$, where $\mathcal{D}^{\mu}=\sqrt{-g}\Delta^{\mu}= h J^{\mu}$ is the densitized dilation flux vector and $h$ is the dilation charge of a particle, produces the above form. Including also the appropriate factor for the above term, our action functional reads
\beq
I = \int d^d x \left[ -\sqrt{-g}\rho(n,s,h) + {J}^\mu\Big( \varphi_{,\mu} + s\theta_{,\mu} + \beta_A \alpha^A{}_{,\mu}  - \frac{1}{4}hQ_\mu \Big)\right]\,. \label{action}
\eeq
The first term is the fluid's energy density that now depends also on the dilation charge and the next two terms are there to ensure particle number conservation and the entropy conservation. These appear in the usual perfect fluid formulation \cite{Brown:1992kc}. The new addition here is the last term which produces a non-vanishing dilation.
Variation with respect to the connection produces (\ref{dil}) while variation with respect to the metric gives the energy-momentum tensor
\beq
T^{\mu\nu}=\rho u^{\mu} u^{\nu}+\left[ n\frac{\partial \rho}{\partial n} - \rho\right](g^{\mu\nu}+u^{\mu}u^{\nu})-\frac{1}{2}g^{\mu\nu}\frac{1}{\sqrt{-g}}\partial_{\alpha}(h J^{\alpha})\,. \label{Tht}
\eeq
We find the usual expression for the pressure,
\beq
p:=n\frac{\partial \rho}{\partial n}- \rho \label{pdef}\,. 
\eeq
In addition, we can write the above energy-momentum tensor in the effective form
\beq
T^{\mu\nu}=\rho_{m}u^{\mu}u^{\nu}+p_{m}h^{\mu\nu} \label{T}
\eeq
where obviously has retained a perfect fluid form.
In the above, the dilation modified density and pressure have been defined through
\beq
\rho_{m}=\rho+\frac{1}{2 \sqrt{-g}}\partial_{\alpha}(\sqrt{-g}nh u^{\alpha})
\eeq
\beq
p_{m}=p-\frac{1}{2 \sqrt{-g}}\partial_{\alpha}(\sqrt{-g}nh u^{\alpha}) \label{ppm}
\eeq
the subscript $m$ denotes that the associated quantity is derived from the metrical energy-momentum tensor.
We see that an incoming (outgoing) dilation flux reduces (enhances) the density and enhances (reduces) the fluid's pressure. Interestingly, if the dilation current is conserved  (i.e. $\partial_{\alpha}(\sqrt{-g}h u^{\alpha})=0$) density and pressure do not receive further modifications. This is the case for  frame rescaling invariant matter \cite{iosifidis2019scale}. Note that, quite interestingly, the combination $\rho+p$ is unaffected by the presence of dilation. For later convenience let us also set
\beq
\mathcal{D}:=\frac{1}{ \sqrt{-g}}\partial_{\alpha}(h J^{\alpha})=\tilde{\nabla}_{\mu}(nhu^{\mu})
\eeq

As a self-consistency check we note that from (\ref{T}) it follows that the total density and pressure $\rho$ and $p$ are the ones associated to the canonical energy-momentum tensor as they should. Indeed, if $t^{\mu\nu}=\rho_{c}u^{\mu}u^{\nu}+p_{c}h^{\mu\nu}$ represents the canonical energy-momentum tensor, from one of the conservation laws of MAG  
\beq
	t{^\mu{}_{\lambda}}=T{^\mu{}_{\lambda}}+\frac{1}{2 \sqrt{-g}}(\nabla_{\nu}-2S_{\nu})(\sqrt{-g}\Delta{_\lambda{}^{\mu\nu}}) \label{conlaw1}
	\eeq
 for pure dilation one easily obtains
 \beq
t^{\mu\nu}=T^{\mu\nu}+\frac{1}{2}g^{\mu\nu} \frac{1}{ \sqrt{-g}}\partial_{\alpha}(h J^{\alpha}) \label{tTrel}
 \eeq
and taking the two projections it follows that $\rho=\rho_{c}$ and $p=p_{c}$, meaning that the density that appears in our functional is the canonical that contains contributions both from $T_{\mu\nu}$ (the usual $\rho_{m}$) and hypermomentum (the divergence part of the dilation). For dilation with vanishing divergence, and only then, the canonical coincides with the metrical as seen from (\ref{tTrel}).

Variations with respect to the rest of the fluid variables, yield
\bs
\ba
\text{particle flux density:} \quad & 0  = & \delta I/\delta J^\mu = \mu u_\mu + \varphi_{,\mu} + s\theta_{,\mu} + \beta_A\alpha^A{}_{,\mu} - \frac{1}{4}hQ_\mu\,, \label{Jcons} \\
\text{chemical free energy potential:} \quad & 0  = & \delta I/\delta \varphi = - \partial_\mu J^\mu\,, \label{ncons} \\
\text{thermacy:} \quad & 0  = & \delta I/\delta \theta  = - \partial_\mu\lp s J^\mu\rp\,, \label{scons} \\ 
\text{entropy per particle:} \quad & 0  = & \delta I/\delta s  = -\sqrt{-g}\frac{\partial\rho}{\partial s} + \theta_{,\mu} J^\mu\,, \\
\text{Lagrangian coordinates:} \quad & 0  = & \delta I/\delta \alpha^A = -\partial_{\mu}(J^{\mu}\beta_{A})\,, \label{acons} \\
\text{The dual coordinates:} \quad & 0  = & \delta I/\delta \beta_A = J^\mu\alpha^A{}_{,\mu}\,, \label{bcons} \\
\text{dilation per particle:} \quad & 0  = & \delta I/\delta h = -\sqrt{-g}\frac{\partial\rho}{\partial h} -\frac{1}{4}\sqrt{-g}n Q_{\mu}u^\mu  \,.   \label{hcons}
\ea
\es
For clarity, we have stated systematically each physical quantity and written explicitly the equation of motion for each of the respective variational degree of freedom. 
Now  we define in the usual manner the chemical potential according to
\beq
\mu:=\frac{\partial \rho}{\partial n}\,,
\eeq
and recall also that the temperature is given by
\beq
T:=\frac{1}{n}\frac{\partial \rho}{\partial s}
\eeq
We may now define a new conjugate variable to dilation:
\beq
\xi:= \frac{1}{n}\frac{\partial \rho}{\partial h}\,.
\eeq
The free energy of the system is then expressed as
\beq
f:=\mu-Ts-h\xi
\eeq
Using the equations of motion it is not difficult to see that
\beq
T=u^{\mu}\partial_{\mu}\theta \label{Ttheta}
\eeq
and
\beq
f=u^{\mu}\partial_{\mu}\phi
\eeq
and therefore, $\theta $ and $\phi$ are potentials for the temperature and the free energy respectively. Let us also note that in the case of an integrable Weyl vector, 
\beq
Q_{\mu}=-4\partial_{\mu}\lambda
\eeq
we would have $\xi=u^{\mu}\partial_{\mu}\lambda$ and the $\lambda$ would serve as a dilaton thermacy.
In our case the first law reads
\beq
d \rho =\mu dn+n T ds+n\xi dh
\eeq
which means that an equation of state of the form $\rho=\rho(n,s, h)$  is provided.
Given definition (\ref{pdef}) for pressure\footnote{This expression with the new definitions reads $ p=n\mu-\rho$.} this can also be written as
\beq
n^{-1}dp=d \mu-Tds - \xi dh\,.  \label{dp}
\eeq
  which suggests an equation of state of the form $p=p(\mu,s,h)$. Now, as in the classical case, the equations of motion should imply particle number conservation and also give the equation of motion of the canonical energy-momentum tensor. The former is explicitly manifested in (\ref{ncons}). As for the latter, in the case of pure dilation it is easy to show that its conservation law becomes
  \beq
\tilde{\nabla}_{\mu}t^{\mu}{}_{\alpha}=\frac{1}{2}\hat{R}_{\nu\alpha}\Delta^{\nu}+\frac{1}{4}Q_{\alpha}\mathcal{D} \label{con2}
  \eeq
With projection along the streamlines
\beq
u^{\alpha}\tilde{\nabla}_{\mu}t^{\mu}{}_{\alpha}=\frac{1}{4}Q_{\mu}u^{\mu}\mathcal{D} \label{proju}
\eeq
We shall now show that this conservation law is implied by our equations of motion as it should. Acting with $u^{\mu}\tilde{\nabla}_{\mu}$ on $t^{\mu}{}_{\alpha}=(\rho+p)u^{\mu}u_{\alpha}+p\delta_{\alpha}^{\mu}$ it follows that
\beq
u^{\alpha}\tilde{\nabla}_{\mu}t^{\mu}{}_{\alpha}=-\Big[ u^{\mu}\partial_{\mu}\rho+(\rho+p)\tilde{\nabla}_{\mu}u^{\mu} \Big]
\eeq
Then plugging
\beq
u^{\mu}\partial_{\mu}\rho=\frac{\partial \rho}{\partial n}u^{\alpha}\partial_{\alpha}n+\frac{\partial \rho}{\partial s}u^{\alpha}\partial_{\alpha}s+\frac{\partial \rho}{\partial h}u^{\alpha}\partial_{\alpha}h
\eeq
in the above we readily find
\beq
-u^{\alpha}\tilde{\nabla}_{\mu}t^{\mu}{}_{\alpha}=\mu \tilde{\nabla}_{\mu}(nu^{\mu}) + nT u^{\mu}\partial_{\mu}s + n\xi  u^{\mu}\partial_{\mu} h \label{ut}
\eeq
where we have also used the fact that
\beq
p+\rho=n \mu\,. \label{prho}
\eeq
Now, the first and second terms on the right-hand side of (\ref{ut}) vanish by virtue of the particle number and entropy conservation laws (\ref{ncons}) and (\ref{scons}) respectively. Finally the last term, once we add a $0=\xi h \tilde{\nabla}_\alpha (nu^\alpha)$, use the definitions for $\xi$ and $\mathcal{D}$ and the dilation equation of motion (\ref{hcons}), becomes
\beq
n\xi  u^{\mu}\partial_{\mu} h + \xi h\tilde{\nabla}_\mu (n u^\mu) =  \xi\tilde{\nabla}_{\mu}(nh u^{\mu})=-\frac{1}{4}Q_{\mu}u^{\mu}\mathcal{D}
\eeq
and further substitution of this into (\ref{ut}) establishes  equivalence with (\ref{proju}) as expected.

Let us now focus on the orthogonal projection of the conservation law (\ref{con2}). The orthogonal projection of its left-hand side reads
\beq
h^\alpha_{\nu}\tilde{\nabla}_{\mu}t^{\mu}{}_{\alpha}=(\rho+p)u^{\mu}\tilde{\nabla}_{\mu}u_{\nu}+u_{\nu}u^{\mu}\partial_{\mu}p+\partial_{\nu}p \label{orthproj}
\eeq
where we have used the fact that $u^{\alpha}\tilde{\nabla}_{\mu}u_{\alpha}=0$ as a consequence of $u^{\alpha}u_{\alpha}=-1$. Then differentiating (\ref{pdef}) it follows that\footnote{This follows also, most easily, from equation (\ref{dp}) by writing the differentials as $dF=d\tau u^{\mu}\partial_{\mu}F$ and comparing left and right-hand sides of the resulting expression.}
\beq
n^{-1}\partial_{\nu}p=\partial_{\nu}\mu -T \partial_{\nu}s  - \xi\partial_{\nu} h\,.
\eeq
Substituting this into the above, using (\ref{prho}) and noting the Leibniz rule in the relevant terms we arrive at
\beq
h^\alpha_{\nu}\tilde{\nabla}_{\mu}t^{\mu}{}_{\alpha}=nu^{\mu}\tilde{\nabla}_{\mu}(\mu u_{\nu}) + n \partial_\nu \mu- nT\lp u_\nu u^{\mu}\tilde{\nabla}_{\mu} s + \partial_\nu s\rp  - n\xi\lp u_\nu u^{\mu}\tilde{\nabla}_{\mu} h + \partial_\nu h\rp\,. 
\eeq
Next, observe that
\beq
\partial_{\nu}\mu=-u^{\mu}\tilde{\nabla}_{\nu}(\mu u_{\mu})
\eeq
as a consequence of $u^{\mu}\tilde{\nabla}_{\nu}u_{\mu}=0$. It follows that
\beq
h^\alpha_{\nu}\tilde{\nabla}_{\mu}t^{\mu}{}_{\alpha}=2 nu^{\mu}\tilde{\nabla}_{[\mu}(\mu u_{\nu]})
- nT \partial_\nu s  - n\xi\lp u_\nu u^{\mu}\tilde{\nabla}_{\mu} h + \partial_\nu h\rp\,, \label{result}
\eeq
where we had recalled the entropy exchange constraint, i.e. $-nT u_{\nu}u^{\mu}\tilde{\nabla}_{\mu}s=0$.  This result we should now compare with the orthogonal projection of (\ref{con2}). It reads, in general,
\beq
h^\alpha_{\nu}\tilde{\nabla}_{\mu}t^{\mu}{}_{\alpha}=\frac{1}{2}\hat{R}_{\mu\nu}\Delta^{\mu}+\frac{\mathcal{D}}{4}\Big( Q_{\nu}+(Q_{\alpha}u^{\alpha})u_{\nu}\Big)
\eeq
and when combined with the result (\ref{result}), in particular
\beq 
2 nu^{\mu}\tilde{\nabla}_{[\mu}(\mu u_{\nu]})
- nT \partial_\nu s  - n\xi\lp u_\nu u^{\mu}\tilde{\nabla}_{\mu} h + \partial_\nu h\rp=\frac{1}{2}\hat{R}_{\mu\nu}\Delta^{\mu}+\frac{\mathcal{D}}{4}\Big( Q_{\nu}+(Q_{\alpha}u^{\alpha})u_{\nu}\Big) \label{ytt}
\eeq
We must now show that this relation is also implied by the equations of motion of our variational formulation. To do so we first take (\ref{Jcons}) act it with $u^{\mu}\tilde{\nabla}_{\mu}$ and antisymmetrize in the lower indices, which results in
\beq
 u^{\mu}\tilde{\nabla}_{[\mu}(\mu u_{\nu]})+u^{\mu}\tilde{\nabla}_{[\mu}\partial_{\nu]}\phi+u^{\mu}(\tilde{\nabla}_{[\mu}s) 
 \partial_{\nu]}\theta+s\tilde{\nabla}_{[\mu} \partial_{\nu]}\theta+\tilde{\nabla}_{[\mu}(\beta_{A}\partial_{\nu]}\alpha^{A})=
 \frac{1}{4}u^\mu\tilde{\nabla}_{[\mu}(hQ_{\nu]})\,.
\eeq
Using then the rest of the equations of motion, in particular (\ref{scons}), (\ref{acons}) and (\ref{bcons}) and the fact that the Levi-Civita covariant derivative is torsionless, it easily follows that
\beq
u^{\mu}\tilde{\nabla}_{[\mu}\partial_{\nu]}\phi=0=u^{\mu}\tilde{\nabla}_{[\mu}\partial_{\nu]}\theta=u^{\mu}\tilde{\nabla}_{[\mu}(\beta_{A}\partial_{\nu]}\alpha^{A})=u^{\mu}(\tilde{\nabla}_{\mu}s) 
 \partial_{\nu}\theta
\eeq
and given also relation (\ref{Ttheta}) for the temperature, we conclude that
\beq
2u^{\mu}\tilde{\nabla}_{[\mu}(\mu u_{\nu]})-T\partial_{\nu}s = \frac{1}{2}u^\mu\tilde{\nabla}_{[\mu}(hQ_{\nu]}) = 
\frac{1}{2}u^\mu\lp h_{,[\mu}Q_{\nu]} + h\hat{R}_{\mu\nu}\rp\,.
\eeq
Quite interestingly, this equation which is fundamental in the perfect fluid formulation \cite{Brown:1992kc}, is modified by geometrical terms due to the presence of dilation. Plugging this back into (\ref{ytt}), we cancel three of the terms in that equation and are left with
\beq
 - n\xi\lp u_\nu u^{\mu}\tilde{\nabla}_{\mu} h + \partial_\nu h\rp = -\frac{1}{2}n u^\mu h_{,[\mu}Q_{\nu]} 
+\frac{\mathcal{D}}{4}\Big( Q_{\nu}+(Q_{\alpha}u^{\alpha})u_{\nu}\Big) \label{ytt2}
\eeq
It remains to show that this is also implied by the equations of motion. In doing so we first use (\ref{hcons}) to substitute the contraction $Q_{\mu}u^{\mu} = -4\xi$ on the right hand side and divide by $n$, 
\beq
 - \xi\lp u_\nu u^{\mu}\tilde{\nabla}_{\mu} h + \partial_\nu h\rp = -\frac{1}{2} u^\mu h_{,[\mu}Q_{\nu]} + (u^\mu\tilde{\nabla}_\mu h)\Big(  \frac{1}{4}Q_\nu -\xi u_\nu\Big) = -\xi(\delta^\mu_\nu + u^\mu u_\nu) h_{,\mu}\,.
\eeq
Of course, the left hand of the equation is neither nothing else than the spatial derivative of the dilaton charge $h$ (times minus the dilatonic potential $\xi$).  Therefore eq. (\ref{orthproj}) is indeed implied by our equations of motion, proving the self-consistency of our variational formulation!

It is quite interesting that the "on-shell" action of our formulation continues to be the integral of just the pressure. Indeed, from the equations of motion it directly follows that on-shell we have that
\beq
S_{on-shell}=\int d^{d}x \sqrt{-g}p
\eeq
Notice that here the full hyperpressure appears but the substitution of (\ref{ppm}) into the latter reveals that this is the same with $\int d^{d}x \sqrt{-g}p_{m}$ since the presence of dilation will simply add a boundary term. Of course, it goes without saying, that the on-shell action is not unique and can take various forms by adding surface integrals to it \cite{Brown:1992kc}.

 It is important to note that instead of (\ref{action}) we could just as well have started from the action
 \beq
I = \int d^d x \left[ -\sqrt{-g}\rho(n,s,h) + {J}^\mu\Big( \varphi_{,\mu} + s\theta_{,\mu} + \beta_A \alpha^A{}_{,\mu}\Big) -\frac{1}{4}Q_{\mu}\mathcal{D}^{\mu} \right] \label{actionee}
\eeq
which is obtained from (\ref{action}) by the mere rescaling $h\rightarrow h/n$. Then the fluid  equations of motion would read
\bs
\ba
\text{particle flux density:} \quad & 0  = & \delta I/\delta J^\mu = \mu u_\mu + \varphi_{,\mu} + s\theta_{,\mu} + \beta_A\alpha^A{}_{,\mu}-\frac{1}{4}\frac{h}{n}Q_{\alpha}h^{\alpha}{}_{\mu} \,, \label{Jcons1} \\
\text{chemical free energy potential:} \quad & 0  = & \delta I/\delta \varphi = - \partial_\mu J^\mu\,,  \\
\text{thermacy:} \quad & 0  = & \delta I/\delta \theta  = - \partial_\mu\lp s J^\mu\rp\,,  \\ 
\text{entropy per particle:} \quad & 0  = & \delta I/\delta s  = -\sqrt{-g}\frac{\partial\rho}{\partial s} + \theta_{,\mu} J^\mu\,, \\
\text{Lagrangian coordinates:} \quad & 0  = & \delta I/\delta \alpha^A = -\partial_{\mu}(J^{\mu}\beta_{A})\,,  \\
\text{The dual coordinates:} \quad & 0  = & \delta I/\delta \beta_A = J^\mu\alpha^A{}_{,\mu}\,,  \\
\text{dilation:} \quad & 0  = & \delta I/\delta h = -\frac{\partial\rho}{\partial h} -\frac{1}{4} Q_{\mu}u^\mu \,.   
\ea
\es
and the associated energy-momentum tensor
\beq
T^{\mu\nu}=\rho_{m}u^{\mu}u^{\nu}+p_{m}h^{\mu\nu} \label{Tttt}
\eeq
where
\beq
\rho_{m}=\rho+\frac{1}{2 \sqrt{-g}}\partial_{\alpha}(\sqrt{-g}h u^{\alpha})
\eeq
\beq
p_{m}=p-\frac{1}{2 \sqrt{-g}}\partial_{\alpha}(\sqrt{-g}h u^{\alpha}) \label{ppm}
\eeq
In the above we have defined the generalized 'hyper-pressure' as
\beq
p:=n\frac{\partial \rho}{\partial n}+ h\frac{\partial \rho}{\partial h}- \rho \label{pdef2}
\eeq
The energy-momentum tensor (\ref{Tttt}) is directly obtained by (\ref{Tht}) by setting $h\rightarrow h/n$ in the latter. Comparing then the two approaches we see here that in the rescaled case (\ref{actionee}) the $u^{\mu}$ projection of the particle flux density equation remains unaltered with respect to its usual perfect fluid form on the cost of changing the corresponding pressure to be given by the hyper-form (\ref{pdef2}) instead of its usual form (\ref{pdef}). On the other hand, in the formulation given by (\ref{action}) the pressure retains its standard definition but the particle flux density equation gets modified, or equivalently the definition of free energy gets modified\footnote{For instance in the pure dilation case of this section, this reads $f=\mu-T s-h\xi$.}. Both approaches describe the same physics and it is a matter of choice which one to use. In the forthcoming discussion we will chose to use the rescaled version and consequently keep the $u^{\mu}$-projected particle flux density equation, or equivalently the definition of free energy, unaltered.  More information about the transition from one set of variables to the other can be found in the appendix.

\subsection{Example: Barotropic Dilation Hyperfluids}

We shall now model our construction, namely we will consider some given functions $\rho(n,h)$ that will reproduce and consequently extend some standard results. Let  us start with the good old barotropic fluid (i.e. derive the equation of state $p=w\rho$), which is very important since it is successfully used in may physical problems. The difference now is that $\rho$ depends also on the dilation charge $h$. This, however, does not cause any difficulty, since by just looking (\ref{pdef})  it is obvious that any homogeneous function $\rho(n,h)$ (in the Euler sense) of some given degree, say $m$ will give the desired barotropic equation of state for the pressure $p$. Indeed, if the density is homogeneous of degree $m$, then by Euler's Theorem for homogeneous functions, if
\beq
\rho(tn,th)=t^{m}\rho(n,h)
\eeq
it follows that
\beq
n\frac{\partial \rho}{\partial n}+ h\frac{\partial \rho}{\partial h}=m \rho
\eeq
and substituting this into the pressure definition (\ref{pdef}) we obtain the usual form of the barotropic equation of state, viz.
\beq
p=w \rho,   \qquad w=m-1
\eeq
as stated. It should be noted however that this equation of state holds true here for the general pressure that appears in the canonical energy-momentum tensor. If $\rho$ was homogeneous only in $n$ then the corresponding barotropic equation of state would hold true only for the perfect fluid component of $\rho$. Let us now  consider the specific example
\beq
\rho(n,h)=Cn^{a}h^{b}
\eeq
It then follows that
\beq
\mu=\frac{\partial \rho}{\partial n}=\frac{a}{n}\rho
\eeq
\beq
\xi=\frac{\partial \rho}{\partial h}=\frac{b}{h}\rho
\eeq
and form (\ref{hcons}) we may solve for the dilation charge as
\beq
h=-\frac{4b}{Q_{\mu}u^{\mu}}\rho
\eeq
and the hyper-pressure is given by
\beq
p=(a+b-1)\rho
\eeq
Notice that if the phenomenological parameters satisfy $a+b=1$ we have a hyper-dust, i.e. $p=0$.

\section{Pure Spin Hyperfluid}
\label{spin}

The next task is to formulate the Action Principle for the pure spin Hyperfluid.  Let us first recall the two spin current densities 
\beq
\sigma^{\mu}:=\sqrt{-g}\sigma u^{\mu}
\eeq
\beq
\zeta^{\mu}:=\sqrt{-g}\zeta u^{\mu}
\eeq
and the pure spin hypermomentum tensor to be obtained from the variational principle reads (see also eq. (\ref{spinD}))
\beq
\Delta_{\alpha\mu\nu}=\Delta_{[\alpha\mu]\nu}=\sigma u_{[\alpha}g_{\mu]\nu}+\epsilon_{\alpha\mu\nu\kappa}u^{\kappa}\zeta \label{sp1}
\eeq
Then, using the variations 
\begin{equation}
	\delta_{\Gamma}S_{\alpha}=\delta_{\Gamma}S_{\alpha\lambda}{}{}^{\lambda}=\delta_{\alpha}^{[\mu}\delta_{\lambda}^{\nu]}\delta\Gamma^{\lambda}{}{}_{\mu\nu}
	\end{equation}
\beq
	\delta_{\Gamma}Q_{\rho}=2 \delta_{\rho}^{\nu}\delta_{\lambda}^{\mu}\delta \Gamma^{\lambda}{}_{\mu\nu}
	\eeq
\begin{equation}
	\delta_{\Gamma}q_{\beta}=(g^{\mu\nu}g_{\beta\lambda}+\delta_{\beta}^{\mu}\delta_{\lambda}^{\nu})\delta \Gamma^{\lambda}{}_{\mu\nu}
	\end{equation}
it can be easily checked that the Lagrangian density whose  $\Gamma$-variation reproduces (\ref{sp1})  is the following 
\beq
\sqrt{-g}\mathcal{L}_{Spin}=-\frac{1}{2}\Big( -2 S_{\mu}+\frac{1}{2}q_{\mu}-\frac{1}{2}Q_{\mu}\Big) \sigma^{\mu}+\frac{1}{2}t_{\mu}\zeta^{\mu}
\eeq
Accordingly, the full pure spin fluid action is given by
\beq
I = \int d^d x \left[ -\sqrt{-g}\rho(n,s,\sigma, \zeta) + {J}^\mu\Big( \varphi_{,\mu} + s\theta_{,\mu} + \beta_A \alpha^A{}_{,\mu}\Big) -\frac{1}{2}\Big( -2 S_{\mu}+\frac{1}{2}q_{\mu}-\frac{1}{2}Q_{\mu}\Big) \sigma^{\mu}+\frac{1}{2}t_{\mu}\zeta^{\mu} \right] \label{actionSpin}
\eeq
Variation with respect to the metric then gives us the energy-momentum tensor associated to  the pure spin Hyperfluid
\begin{gather}
T_{\mu\nu}=\Big[ \rho-\frac{1}{2\sqrt{-g}}(\partial_{\alpha}\sigma^{\alpha}+\zeta^{\alpha}t_{\alpha})\Big]u_{\mu}u_{\nu}+\Big[ p+\frac{1}{2\sqrt{-g}}(\partial_{\alpha}\sigma^{\alpha}+\zeta^{\alpha}t_{\alpha})\Big]h_{\mu\nu} \nonumber \\
+\frac{1}{2}\Big[ \sigma A_{(\mu}u_{\nu)}-\sigma \nabla_{(\mu}u_{\nu)}-u_{(\mu}\partial_{\nu)}\sigma \Big]-2\zeta u^{\rho}\epsilon_{\rho\alpha\beta(\mu}S_{\nu)}{}^{\alpha\beta} \label{Tmnspin}
\end{gather}
where $A_{\mu}:=2S_{\mu}-q_{\mu}+\frac{1}{2}Q_{\mu}$ and the spin influenced hyper-pressure is now defined as
\beq
p:=\sigma \frac{\partial \rho}{\partial \sigma}+\zeta \frac{\partial \rho}{\partial \zeta}-\rho
\eeq
It is quite remarkable that the above energy-momentum tensor describes, in fact, not a perfect fluid sector but rather an imperfect one. Indeed, let us recall that the energy-momentum tensor of a generic imperfect fluid has the form
\beq
T_{\mu\nu}=\hat{\rho} u_{\mu}u_{\nu}+\hat{p} h_{\mu\nu}+2\mathcal{Q}_{(\mu}u_{\nu)}+\pi_{\mu\nu}
\eeq
where as usual the energy density and the isotropic pressure of the fluid are given by\footnote{We use hats here in order to denote the net value quantities.}
\beq
\hat{\rho}:=u^{\mu}u^{\nu}T_{\mu\nu}
\eeq
and
\beq
\hat{p}:=\frac{1}{(d-1)}h^{\mu\nu}T_{\mu\nu}
\eeq
respectively. In addition, we now have the heat flow vector (or heat flux)
\beq
\mathcal{Q}_{\alpha}:=-u^{\mu}h^{\nu}{}_{\alpha}T_{\mu\nu}
\eeq
along with the anisotropic stresses
\beq
\pi_{\alpha\beta}:=h^{\mu}{}_{\alpha}h^{\nu}{}_{\beta}T_{\mu\nu}-\frac{1}{(d-1)}(T_{\mu\nu}h^{\mu\nu})h_{\alpha\beta}=h^{\mu}{}_{\alpha}h^{\nu}{}_{\beta}T_{\mu\nu}-\hat{p} h_{\alpha\beta}
\eeq
Taking the above definitions into consideration, for the pure spin fluid energy-momentum tensor (\ref{Tmnspin}) we easily compute\footnote{Notice the appearance of a term $\propto \tilde{\Theta}$ also in the expression for density (apart from the pressure), in contrast to the Eckart model and in accordance with the EFT appoach to relativistic hydrodynamics. This term appears also in Horndeski Theories with imperfect fluids (see \cite{pujolas2011imperfect}).}
\beq
\hat{\rho}=\rho -\frac{1}{2 \sqrt{-g}}\zeta^{\alpha}t_{\alpha}+\frac{\sigma}{2}\Big[ -\tilde{\Theta}+\frac{1}{2}Q_{\alpha\beta\gamma}u^{\alpha}u^{\beta}u^{\gamma}-A_{\mu}u^{\mu} \Big] \label{visrho}
\eeq
\beq
(d-1)\hat{p}=(d-1)p+\frac{(d-1)}{2}u^{\alpha}\partial_{\alpha}\sigma+\frac{\sigma}{2}(d-2)\tilde{\Theta}+\frac{\sigma}{2}\Big[\frac{1}{2}Q_{\alpha\beta\gamma}u^{\alpha}u^{\beta}u^{\gamma}-A_{\mu}u^{\mu} \Big]- \frac{1}{2 \sqrt{-g}}\zeta^{\alpha}t_{\alpha} \label{visp}
\eeq
where $\tilde{\Theta}:=\tilde{\nabla}_{\mu}u^{\mu}$ is the usual Riemannian expansion scalar. As for the heat flux we find
\begin{gather}
\mathcal{Q}_{\lambda}=\frac{\sigma}{4}\Big[ A_{\lambda}+a_{\lambda}-\frac{\partial_{\lambda}\sigma}{\sigma}-\frac{1}{2}Q_{\lambda\alpha\beta}u^{\alpha}u^{\beta} \Big]+\frac{\sigma}{4}\Big[ A_{\mu}u^{\mu}-u^{\mu}\frac{\partial_{\mu}\sigma}{\sigma}-\frac{1}{2}Q_{\alpha\beta\gamma}u^{\alpha}u^{\beta}u^{\gamma} +a_{\mu}u^{\mu}\Big]u_{\lambda}+\zeta u^{\rho}\epsilon_{\rho\alpha\beta\lambda}S_{\mu}{}^{\alpha\beta}u^{\mu}
\end{gather}
or in fully expanded form
\begin{gather}
\mathcal{Q}_{\lambda}=\frac{\sigma}{4}\Big[ A_{\lambda}+\tilde{a}_{\lambda}-\frac{\partial_{\lambda}\sigma}{\sigma}-(Q_{\lambda\alpha\beta}+2 S_{\lambda\alpha\beta})u^{\alpha}u^{\beta} \Big]+\frac{\sigma}{4}\Big[ A_{\mu}u^{\mu}-u^{\mu}\frac{\partial_{\mu}\sigma}{\sigma}-Q_{\alpha\beta\gamma}u^{\alpha}u^{\beta}u^{\gamma} \Big]u_{\lambda}+\zeta u^{\rho}\epsilon_{\rho\alpha\beta\lambda}S_{\mu}{}^{\alpha\beta}u^{\mu}
\end{gather}
and for the stress tensor 
\begin{gather}
    \pi_{\alpha\beta}=-\frac{\sigma}{2}\left[ \nabla_{(\alpha}u_{\beta)}-\frac{1}{2}u_{(\alpha}Q_{\beta)\mu\nu}u^{\mu}u^{\nu}+u_{(\alpha}a_{\beta)}-\frac{1}{2}Q_{\mu\nu\lambda}u^{\mu}u^{\nu}u^{\lambda}u_{\alpha}u_{\beta} \right]-2\zeta u^{\rho}\epsilon_{\rho\mu\nu(\alpha}\Big( S_{\beta)}{}^{\mu\nu}+u_{\beta)}S_{\lambda}{}^{\mu\nu}u^{\lambda} \Big) \nonumber \\
    +\Big(p+ \frac{1}{2\sqrt{-g}}\partial_{\alpha}\sigma^{\alpha}+ \frac{1}{2 \sqrt{-g}}\zeta^{\alpha}t_{\alpha}\Big)h_{\alpha\beta}-\hat{p}h_{\alpha\beta}
\end{gather}
or
\begin{gather}
  \pi_{\alpha\beta}  =-\frac{\sigma}{2}\left[ \nabla_{(\alpha}u_{\beta)}-\frac{1}{2}u_{(\alpha}Q_{\beta)\mu\nu}u^{\mu}u^{\nu}+u_{(\alpha}a_{\beta)}-\frac{1}{2}Q_{\mu\nu\lambda}u^{\mu}u^{\nu}u^{\lambda}u_{\alpha}u_{\beta} \right]-2\zeta u^{\rho}\epsilon_{\rho\mu\nu(\alpha}\Big( S_{\beta)}{}^{\mu\nu}+u_{\beta)}S_{\lambda}{}^{\mu\nu}u_{\lambda} \Big) \nonumber \\+\frac{1}{2 (d-1)}h_{\alpha\beta}\left[ \frac{4}{ \sqrt{-g}}\zeta^{\alpha}t_{\alpha} +\sigma \Big( \tilde{\Theta}+A_{\mu}u^{\mu} -\frac{1}{2}Q_{\lambda\mu\nu}u^{\lambda}u^{\mu}u^{\nu}\Big) \right]
\end{gather}
 It is an easy matter to show that this is indeed traceless ($\pi_{\alpha}{}^{\alpha}=0$) an orthogonal to $u^{\mu}$ (i.e. $\pi_{\alpha\beta}u^{\beta}=0$) as it should.
However, this is not the end of the story. If we perform a post-Riemannian expansion on $\nabla_{\alpha}u_{\mu}$ and also use the definition
\beq
\tilde{\sigma}_{\mu\nu}:=h^{\alpha}_{(\mu}h^{\beta}_{\nu)}\tilde{\nabla}_{\alpha}u_{\beta}-\frac{\tilde{\Theta}}{(d-1)}h_{\mu\nu}
\eeq
for the usual Riemannian kinematic shear $\tilde{\sigma}_{\mu\nu}$ , or the more general (Riemannian) split
\beq
\tilde{\nabla}_{\beta}u_{\alpha}=\tilde{\sigma}_{\alpha\beta}+\frac{\tilde{\Theta}}{(d-1)}h_{\alpha\beta}+\omega_{\alpha\beta}-\tilde{a}_{\alpha}u_{\beta}
\eeq
then the above expression for the anisotropic stresses takes the form
\begin{gather}
    \pi_{\alpha\beta}  =-\frac{\sigma}{2}\left[ \tilde{\sigma}_{\alpha\beta}-N_{\lambda(\alpha\beta)}u^{\lambda}-u_{(\alpha}N_{\mu|\beta)\nu}u^{\mu}u^{\nu}-\frac{1}{2}u_{(\alpha}Q_{\beta)\mu\nu}u^{\mu}u^{\nu}-\frac{1}{2}Q_{\mu\nu\lambda}u^{\mu}u^{\nu}u^{\lambda}u_{\alpha}u_{\beta} \right]\nonumber \\-2 \zeta u^{\rho}\epsilon_{\rho\mu\nu(\alpha}\Big( S_{\beta)}{}^{\mu\nu}+u_{\beta)}S_{\lambda}{}^{\mu\nu}u^{\lambda} \Big) +\frac{1}{(d-1)}h_{\alpha\beta}\left[ \frac{2}{ \sqrt{-g}}\zeta^{\alpha}t_{\alpha} +\frac{\sigma}{2} \Big( A_{\mu}u^{\mu} -\frac{1}{2}Q_{\lambda\beta\gamma}u^{\lambda}u^{\beta}u^{\gamma}\Big) \right]
\end{gather}
Note that various cancellations took place that in the end rendered the expression for the stresses independent of the acceleration $\tilde{a}^{\mu}$ and of the expansion $\tilde{\Theta}$. Now what is quite remarkable   about this final expression is the presence of the term
\beq
\pi_{\alpha\beta} \supseteq -\frac{\sigma}{2}\tilde{\sigma}_{\alpha\beta}
\eeq
in the form of the stress tensor. Let us also observe the appearance of 
\beq
\hat{\rho}\supseteq -\frac{\sigma}{2} \tilde{\Theta}
\eeq
as well as
\beq
\hat{p}\supseteq \frac{(d-2)}{2}\sigma \tilde{\Theta}
\eeq
in the expressions for the viscous density and pressure, see respectively equations (\ref{visrho}) and (\ref{visp}).
In the development of relativistic hydrodynamics, the above expressions are imposed  as constitutive relations by demanding compliance with the second law of thermodynamics. The remarkable fact here is that in our formulation these constitutive relations were not imposed but rather naturally appeared dynamically  due to the inclusion of a spin part of hypermomentum! What is more, the coefficients of bulk and shear viscosity are not free functions but rather they are explicitly given in terms of the spin variables.

Furthermore, from equation (\ref{conlaw1}) we compute the canonical energy momentum tensor to be
\begin{gather}
   t_{\mu\nu}=T_{\mu\nu}-\frac{\sigma}{4}\left[ 2A_{[\mu}u_{\nu]}-2 \frac{u_{[\nu}\partial_{\mu]}\sigma}{\sigma}-q_{\nu}u_{\mu}-2\tilde{a}_{[\mu}u_{\nu]}+(2 S_{\nu\mu\alpha}+Q_{\nu\mu\alpha})u^{\alpha} \right]-\frac{\sigma}{2}\tilde{\omega}_{\mu\nu} \nonumber \\
   +\frac{1}{2}\epsilon_{\alpha\beta\gamma\mu}Q^{\alpha\beta}{}_{\nu}u^{\gamma}\zeta+\frac{1}{2}\epsilon_{\mu\nu\beta\gamma}u^{\gamma}\zeta \Big( 2 S^{\beta}-\tilde{a}^{\beta}-\frac{\partial^{\beta}\zeta}{\zeta}\Big) +\frac{1}{2}\zeta\epsilon_{\mu\nu\beta\gamma}\tilde{\omega}^{\beta\gamma}
\end{gather}
It is then worth mentioning that in the canonical tensor, vorticity contributions (i.e. terms proportional to $\tilde{\omega}_{\mu\nu}$) also appear naturally. This is not so surprising if one recalls that in the pure spin case the hypermomentum tensor is antisymmetric in its first two indices. Nevertheless, it is quite remarkable that all kinematic variables, namely $\tilde{\Theta},\tilde{\sigma}_{\mu\nu}, \tilde{\omega}_{\mu\nu}$ and $ \tilde{a}_{\mu}$ appear in the energy tensors. This is not the case in the usual relativistic fluid dynamics because then only the metrical energy-momentum tensor appears which is symmetric by default.
Notice also that the traces are equal, that is $t=T$, as they should given the antisymmetry of the hypermomentum in this case. For completeness let us also report the post-Riemannian expanded version of (\ref{Tmnspin}), which reads
\begin{gather}
    T_{\mu\nu}=\left[ \rho-\frac{u^{\alpha}\partial_{\alpha}\sigma}{2}-\frac{\sigma}{2}\tilde{\Theta}-\frac{1}{2}\zeta t_{\alpha}u^{\alpha} \right]u_{\mu}u_{\nu}+\left[ p+\frac{u^{\alpha}\partial_{\alpha}\sigma}{2}+\frac{1}{2}\zeta t_{\alpha}u^{\alpha} +\frac{\sigma}{2}\frac{(d-2)}{(d-1)}\tilde{\Theta} \right] h_{\mu\nu} \nonumber \\
    -\frac{\sigma}{2} \tilde{\sigma}_{\mu\nu} +\frac{\sigma}{2}u_{(\mu}\Big( A_{\nu)}+\tilde{a}_{\nu)}-\frac{\partial_{\nu)}\sigma}{\sigma} \Big)+\frac{\sigma}{2}N_{\alpha(\mu\nu)}u^{\alpha}-2 \zeta u^{\rho}\epsilon_{\rho\alpha\beta(\mu}S_{\nu)}{}^{\alpha\beta}
\end{gather}

To complete the analysis let us vary (\ref{actionSpin}) with respect to the rest of the fluid variables to arrive at the system of equations
\bs
\ba
0 & = & \delta I/\delta J^\mu = \mu u_\mu + \varphi_{,\mu} + s\theta_{,\mu} + \beta_A\alpha^A{}_{,\mu}+\frac{1}{n}\Big( \sigma \xi_{\alpha}+\frac{\zeta}{2}t_{\alpha}\Big)h^{\alpha}{}_{\mu} \,,  \\
0 & = & \delta I/\delta \varphi = - \mathring{\nabla}_\mu J^\mu\,,  \\
0 & = & \delta I/\delta \theta  = - \mathring{\nabla}_\mu\lp s J^\mu\rp\,, \\ 
0 & = & \delta I/\delta s  = -\sqrt{-g}\frac{\partial\rho}{\partial s} + \theta_{,\mu} J^\mu\,, \\
0 & = & \delta I/\delta \alpha^A = -\partial_{\mu}(J^{\mu}\beta_{A})\,, \\
0 & = & \delta I/\delta \beta_A = J^\mu\alpha^A{}_{,\mu}\,,  \\
0 & = & \delta I/\delta \sigma = -\frac{\partial\rho}{\partial \sigma} +\xi_{\mu} u^{\mu}\,, \label{Sss1}
\\
0 & = & \delta I/\delta \zeta=-\frac{\partial\rho}{\partial \zeta} +\frac{1}{2} t_{\mu}u^{\mu} \,.  \label{Sss2} 
\ea
\es
where
\beq
\xi_{\alpha}:=-\frac{1}{2}\Big( -2 S_{\mu}+\frac{1}{2}q_{\mu}-\frac{1}{2}Q_{\mu}\Big)  \label{XX}
\eeq
The first five of the above remain the same with the case of pure dilation (and also the usual perfect fluid formulation \cite{Brown:1992kc}) and of course the last two come from the spin part.

Having developed the case of the pure dilation and pure spin parts, it now remains to formulate the pure shear case as well in order to complete the analysis of the Hyperfluids. However, let us digress for a little while to discuss general matter-connection couplings of Hyperfluids.

\section{General Couplings to the Connection}
\label{shear}

The particular form of the hypermomentum  we have derived related to the perfect Hyperfluid can be seen as a special case of a universal coupling of matter to the connection. To be more precise, recall that in Electromagnetism, the charge interaction to an external source is described by the interaction Lagrangian
\beq
\mathcal{L}_{int}\propto A_{\mu}J^{\mu} \label{Aint}
\eeq
namely a coupling between the gauge potential and the 4-current source. Then its variation with respect to the gauge potential gives the source term on the right-hand side of the first set of Maxwell equations. Following the Electromagnetism paradigm, we may then write an interaction term between the affine connection and the (now) hypermomentum source. Such an interaction Lagrangian would typically read
\beq
\mathcal{\hat{L}}_{Hyp}=\Gamma^\lambda{}_{\mu\nu}\Delta_\lambda{}^{\mu\nu}
\eeq
However, there is a striking problem with such Lagrangian; obviously it is not covariant given the non-tensorial nature of the affine connection. The task is then to build tensorial objects out of the connection, namely torsion and non-metricity combinations, write down an interaction action of the form
\beq
\mathcal{L}_{Hyp}=\Xi^\lambda{}_{\mu\nu}\Delta_\lambda{}^{\mu\nu}
\eeq
and choose the tensor $\Xi_{\alpha\mu\nu}$ appropriately such that
\beq
-\frac{2}{\sqrt{-g}}\frac{\delta S_{Hyp}}{\delta \Gamma^\lambda{}_{\mu\nu}}=\Delta_\lambda{}^{\mu\nu} \label{Lhyp}
\eeq
It is not so difficult to see that the combination that does the job is given by
\begin{equation}
    \Xi^\lambda{}_{\mu\nu} = \frac{1}{2} g^{\lambda\alpha}\left(Q_{\mu\nu\alpha} + Q_{\nu\mu\alpha}
- Q_{\alpha\mu\nu}\right)+g^{\lambda\alpha}\left( S_{\mu\nu\alpha} - S_{\alpha\mu\nu} - S_{\alpha\nu\mu}\right) \label{Xi}
\end{equation}
and therefore the interaction Lagrangian reads\footnote{Of course the term $-1/2$ is included to compensate for the $-2$ factor in the definition of the Hypermomentum tensor.}
\beq
\mathcal{L}_{Hyp}=-\frac{1}{2}\Big[ \frac{1}{2} g^{\lambda\alpha}\left(Q_{\mu\nu\alpha} + Q_{\nu\mu\alpha}
- Q_{\alpha\mu\nu}\right)+g^{\lambda\alpha}\left( S_{\mu\nu\alpha} - S_{\alpha\mu\nu} - S_{\alpha\nu\mu}\right)  \Big]\Delta_\lambda{}^{\mu\nu} \label{LHD}
\eeq
and is the Hypermomentum analogue of (\ref{Aint}). The observant reader may have noticed that the above combinations of torsion and non-metricity appearing in (\ref{Xi}) constitute exactly the distortion tensor as was defined in (\ref{distortion}). Of course this does not come as a surprise since given an affine connection and the Levi-Civita connection by taking their difference one can form a tensor which can then couple to matter and produce a covariant type interaction. This tensor is non other from the distortion which when varied with respect to the affine connection produces the identity map needed to reproduce $\Delta_\lambda{}^{\mu\nu}$ in (\ref{Lhyp}). So finally the connection-matter interaction is described by
\beq
S_{Hyp}=\int d^{d}x \sqrt{-g}\mathcal{L}_{Hyp}=\int d^{d}x\Big(-\frac{1}{2}\sqrt{-g}N^\lambda{}_{\mu\nu}\Delta_\lambda{}^{\mu\nu} \Big) \label{SmatterH}
\eeq

To double-check this result and reinforce the validity of our statement we may consider purely dilation matter as an example, then given the hypermomentum form (\ref{dil}) one sees that (\ref{LHD}) reduces to
\beq
\mathcal{L}_{Dil}=-\frac{1}{4}Q_{\nu}\Delta^{\nu}
\eeq
which is exactly the extra piece we added into our fluid Lagrangian (\ref{action}) when we constructed the variational principle of the Perfect Dilatonic Hyperfluid. Also, using the spin part (\ref{spinD}), from the above expression we get the interaction Lagrangian density for the pure spin Hyperfluid
\beq
\sqrt{-g}\mathcal{L}_{Spin}=-\frac{1}{2}\Big( -2 S_{\mu}+\frac{1}{2}q_{\mu}-\frac{1}{2}Q_{\mu}\Big) \sigma^{\mu}+\frac{1}{2}t_{\mu}\zeta^{\mu}
\eeq
which exactly matches (\ref{actionSpin}).  Therefore, the interaction term (\ref{LHD}) gives indeed the correct coupling between matter and connection! We shall now use this fact to construct the Lagrangian of a pure shear Hyperfluid. 

\subsection{Pure Shear Hyperfluid Action}

As we already mentioned, given the fact that in the shear part we have a tensorial mode (the one $\propto u_{\alpha}u_{\mu}u_{\nu}$), the construction of the interaction Lagrangian for the pure shear fluid is not as trivial as its pure dilation and pure spin counterparts, where only vectorial modes where excited.  However, with our last observation regarding the form of the interaction Lagrangian (\ref{SmatterH}), we can now easily find the piece that we need to add to the perfect fluid Lagrangian in order to obtain the full pure shear mode. Given the form of the shear mode (\ref{shearDD}) and the interaction Lagrangian density (\ref{SmatterH}) we easily compute
\beq
\sqrt{-g}\mathcal{L}_{shear}=-\frac{1}{2}(\Sigma^{\alpha}+\Pi^{\alpha})N_{\alpha\mu\nu}u^{\mu}u^{\nu}-\frac{1}{2d}\Sigma^{\mu}N_{\mu}^{(1)}-\frac{1}{4}\Pi^{\mu}(N_{\mu}^{(2)}+N_{\mu}^{(3)}) \label{intpshear}
\eeq
and by using the relations between the distortion and torsion and non-metricity we may bring it to a somewhat more physical final form
\beq
\sqrt{-g}\mathcal{L}_{shear}=-\frac{1}{4}(\Sigma^{\alpha}+\Pi^{\alpha})Q_{\alpha\mu\nu}u^{\mu}u^{\nu}-\frac{1}{4d}\Sigma^{\mu}Q_{\mu}-\frac{1}{4}\Pi^{\mu}q_{\mu} \label{intpshear2}
\eeq

which should give the correct addition to reproduce the pure shear effects. It is worth stressing out that there is no appearance of torsion couplings. Furthermore, we see that the sum of the two shear parts couples to the tensor mode of non-metricity while each of the shear vector densities takes one of the two non-metricity vectors.

Computing the associated Hypermomentum tensor for the above Lagrangian density, we find exactly the pure shear form (\ref{shearDD}), which indeed proves the validity of the additional piece (\ref{intpshear}). As a result, the pure shear hyperfluid action reads
\beq
I = \int d^d x \left[ -\sqrt{-g}\rho(n,s,\sigma, \zeta) + {J}^\mu\Big( \varphi_{,\mu} + s\theta_{,\mu} + \beta_A \alpha^A{}_{,\mu}\Big)-\frac{1}{4}(\Sigma^{\alpha}+\Pi^{\alpha})Q_{\alpha\mu\nu}u^{\mu}u^{\nu}-\frac{1}{4d}\Sigma^{\mu}Q_{\mu}-\frac{1}{4}\Pi^{\mu}q_{\mu} \right] \label{actionShear}
\eeq

The associated energy-momentum tensor for the pure-shear Hyperfluid is then computed to be
\begin{gather}
T_{\mu\nu}=\Big[ \rho+\frac{1}{2 d\sqrt{-g}}\partial_{\alpha}\Sigma^{\alpha}+\frac{2}{\sqrt{-g}}\partial_{\alpha}\Xi^{\alpha}+2\frac{\Xi^{\alpha}}{\sqrt{-g}}u^{\beta}u^{\gamma}Q_{\alpha\beta\gamma}\Big]u_{\mu}u_{\nu}+\Big[ p-\frac{1}{2 d\sqrt{-g}}\partial_{\alpha}\Sigma^{\alpha}\Big]h_{\mu\nu} \nonumber \\
+\frac{4}{\sqrt{-g}}\Xi_{(\mu}a_{\nu)}+\frac{4}{\sqrt{-g}}\Xi^{\alpha}Q_{\alpha\beta(\mu}u_{\nu)}u^{\beta}
+\frac{1}{2}\Big[ \Pi A_{(\mu}u_{\nu)}-\Pi \nabla_{(\mu}u_{\nu)}-u_{(\mu}\partial_{\nu)}\Pi \Big]
\end{gather}
where again $A_{\mu}:=2S_{\mu}-q_{\mu}+\frac{1}{2}Q_{\mu}$, we have abbreviated
\beq
\Xi^{\alpha}=-\frac{1}{4}(\Sigma^{\alpha}+\Pi^{\alpha})
\eeq
A  more physical picture may be derived after performing a post-Riemannian expansion,
\begin{gather}
T_{\mu\nu}=\Big[ \rho+\frac{(1-d)}{2 d\sqrt{-g}}\partial_{\alpha}\Sigma^{\alpha}-\frac{1}{2  \sqrt{-g}}\partial_{\alpha}\Pi^{\alpha}-\frac{\Sigma +\Pi}{2}Q_{\alpha\beta\gamma}u^{\alpha}u^{\beta}u^{\gamma}\Big]u_{\mu}u_{\nu}+\Big[ p-\frac{1}{2 d\sqrt{-g}}\partial_{\alpha}\Sigma^{\alpha} -\frac{\Pi}{2(d-1)}\tilde{\Theta} \Big]h_{\mu\nu} \nonumber \\
-\frac{1}{2}(2 \Sigma +\Pi ) \tilde{a}_{(\mu}u_{\nu)}+\frac{1}{2}u_{(\mu}(\Pi A_{\nu)}-\partial_{\nu)}\Pi)-\frac{\Pi}{2}\tilde{\sigma}_{\mu\nu} 
+\frac{\Pi}{2}N^{\lambda}{}_{(\mu\nu)}u_{\lambda}+(\Sigma+\Pi)u^{\alpha}u^{\beta}(N_{\alpha(\mu|\beta|}u_{\nu)}-Q_{\alpha\beta(\mu}u_{\nu)}) \label{Tmnshear}
\end{gather}
where again we recognise the viscous phenomena contributions, with the associated viscous density and pressure given by
\beq
\hat{\rho}=\rho +\frac{(1-d)}{2 d\sqrt{-g}}\partial_{\alpha}\Sigma^{\alpha}-\frac{1}{2  \sqrt{-g}}\partial_{\alpha}\Pi^{\alpha} +\frac{\Pi}{4}Q_{\alpha\mu\nu}u^{\alpha}u^{\mu}u^{\nu}-\frac{\Pi}{2}A_{\mu}u^{\mu}+\frac{1}{2}u^{\mu}\partial_{\mu}\Pi 
\eeq
\beq
(d-1) \hat{p}=(d-1)\Big[ p-\frac{1}{2 d\sqrt{-g}}\partial_{\alpha}\Sigma^{\alpha} -\frac{\Pi}{2(d-1)}\tilde{\Theta} \Big]-\frac{\Pi}{2}A_{\mu}u^{\mu}+\frac{\Pi}{4}Q_{\alpha\mu\nu}u^{\alpha}u^{\mu}u^{\nu}
\eeq

 Expanding further the derivatives in order to separate out the viscous terms we find
 \begin{align}
\hat{\rho}=\rho+\frac{1}{2}\Big[ \frac{(1-d)}{d}\Sigma -\Pi \Big] \tilde{\Theta}+\frac{(1-d)}{2d}u^{\mu}\partial_{\mu}\Sigma   -\frac{\Pi}{2}A_{\mu}u^{\mu}+\frac{\Pi}{4}Q_{\alpha\mu\nu}u^{\alpha}u^{\mu}u^{\nu} \\
(d-1)\hat{p}=(d-1) p+\frac{1}{2}\Big[ \frac{(1-d)}{d}\Sigma -\Pi \Big] \tilde{\Theta}+\frac{(1-d)}{2d}u^{\mu}\partial_{\mu}\Sigma  -\frac{\Pi}{2}A_{\mu}u^{\mu}+\frac{\Pi}{4}Q_{\alpha\mu\nu}u^{\alpha}u^{\mu}u^{\nu}
 \end{align}
from which we read-off the bulk viscosity coefficient\footnote{We use the suffix $visc.$ in the $\zeta$ to distinguish it from the $\zeta$-part of the spin component of hypermomentum.}
\beq
\zeta^{(b)}_{visc.}=\frac{1}{2}\Big[ \Pi-\frac{(1-d)}{d}\Sigma  \Big] 
\eeq
Notice also the  relation
\beq
(d-1)(\hat{p}-p)=(\hat{\rho}-\rho)
\eeq
which is consequence of the vanishing dilation. Given the fact that $\zeta^{(b)}_{visc.}$ is always positive we deduce the constraint
\beq
\Pi>\frac{(1-d)}{d}\Sigma
\eeq
among the shear parts.  As for the heat flux and anisotropic stresses, after a little algebra we find, respectively,
\begin{gather}
\mathcal{Q}_{\lambda}=-\frac{(\Sigma +\Pi)}{2}\tilde{a}_{\lambda}+\frac{\Pi}{4}\Big( A_{\lambda}+\tilde{a}_{\lambda}-\frac{\partial_{\lambda}\Pi}{\Pi}\Big)+\frac{\Pi}{4}\Big( A_{\mu}u^{\mu}-\frac{u^{\mu}\partial_{\mu}\Pi}{\Pi}-Q_{\alpha\beta\gamma}u^{\alpha}u^{\beta}u^{\gamma}\Big)u_{\lambda} \nonumber \\
-\frac{\Pi}{2}u^{\alpha}u^{\beta}N_{\alpha(\beta\lambda)}+\frac{(\Sigma+\Pi)}{2}(N_{\alpha\lambda\beta}u^{\alpha}u^{\beta}-Q_{\alpha\beta\lambda}u^{\alpha}u^{\beta})-\frac{(\Sigma+\Pi)}{4}Q_{\alpha\beta\gamma}u^{\alpha}u^{\beta}u^{\gamma}u_{\lambda}
\end{gather}
\begin{gather}
\pi_{\alpha\beta}=-\frac{\Pi}{2}\tilde{\sigma}_{\alpha\beta}+\frac{\Pi}{2}\left[ u^{\lambda}N_{\lambda(\alpha\beta)}+u^{\lambda}N_{\lambda(\alpha\mu)}u^{\mu}u_{\beta}+u^{\lambda}N_{\lambda(\beta\mu)}u^{\mu}u_{\alpha}+\frac{1}{2}(Q_{\lambda\mu\nu}u^{\lambda}u^{\mu}u^{\nu})u_{\alpha}u_{\beta} \right] \nonumber \\
+h_{\alpha\beta}\frac{\Pi}{2 (d-1)}\Big[ A_{\mu}u^{\mu}-\frac{1}{2}Q_{\lambda\mu\nu}u^{\lambda}u^{\mu}u^{\nu} \Big]
\end{gather}
Note that a shear viscosity coefficient $\zeta^{(s)}_{visk}=-\frac{\Pi}{2}$ can be read-off from the latter equation for the anisotropic stresses.

With the above expression for the energy-momentum tensor and given the relation (\ref{cc1}) and the form of pure shear hypermomentum, we find the following expression for the canonical energy-momentum tensor for the pure shear hyperfluid:

\begin{gather}
    t_{\mu\nu}=T_{\mu\nu}+\frac{1}{2 d \sqg}(\partial_{\alpha}\Sigma^{\alpha})g_{\mu\nu}+\frac{1}{2 \sqg}\Big( \partial_{\alpha}\Sigma^{\alpha}+\partial_{\alpha}\Pi^{\alpha}\Big) u_{\mu} u_{\nu}+\frac{\Pi}{2(d-1)}\tilde{\Theta}(g_{\mu\nu}+u_{\mu}u_{\nu})+\frac{\Pi}{2}\tilde{\sigma}_{\mu\nu}   +\frac{(2 \Sigma +\Pi)}{2}\tilde{a}_{(\mu}u_{\nu)}\nonumber \\
  +\frac{1}{2}u_{(\mu}\partial_{\nu)}\Pi -\frac{1}{2}\Pi A_{(\mu}u_{\nu)}-\frac{1}{4}\Pi u_{\mu}q_{\nu} 
    -\frac{\Pi}{4}(N_{\alpha\nu\mu}-N_{\mu\alpha\nu})u^{\alpha}-\frac{(\Sigma + \Pi)}{2}(N_{\alpha\nu\beta}u_{\mu}-N_{\mu\alpha\beta}u_{\nu})u^{\alpha}u^{\beta}
\end{gather}
As a cross-check, taking the trace of the latter equation, we easily confirm that $t=T$ as we should given that the shear part of the hypermomentum is symmetric and traceless in its two first indices. Furthermore, substituting eq. (\ref{Tmnshear}) in the above relation, various cancellations take place and in the end we arrive at the handy  form
\beq
t_{\mu\nu}=\Big( \rho-\frac{(\Sigma +\Pi)}{2}Q_{\alpha\beta\gamma}u^{\alpha}u^{\beta}u^{\gamma}\Big) u_{\mu}u_{\nu}+p h_{\mu\nu}-\frac{(\Sigma +\Pi)}{2}Q_{\alpha\beta\nu}u^{\alpha}u^{\beta}u_{\mu}+\frac{\Pi}{4}Q_{\nu\mu\alpha}u^{\alpha}-\frac{\Pi}{4} u_{\mu}q_{\nu} \label{cansh}
\eeq
where we see that the density only gets modified by a totally symmetric non-metricity part and the pressure does not change at all. If we define the momentum of the hyperfluid in the standard way (see for instance \cite{deBoer:2017ing}),
\beq
P_{\nu}:=-u^{\mu}t_{\mu\nu}
\eeq
for the above form of the canonical energy-momentum tensor we find the associated total momentum density 
\beq
P_{\nu}=\Big( \rho-\frac{(\Sigma +\Pi)}{2}Q_{\alpha\beta\gamma}u^{\alpha}u^{\beta}u^{\gamma}\Big) u_{\nu}-\frac{(\Sigma + \Pi)}{2}u^{\alpha}u^{\beta}Q_{\alpha\beta\nu}-\frac{\Pi}{4}Q_{\nu\alpha\beta}u^{\alpha}u^{\beta}  -\frac{\Pi}{4}q_{\nu}
\eeq
and with this we may recast (\ref{cansh}) as follows
\beq
t_{\mu\nu}=u_{\mu}P_{\nu}+p h_{\mu\nu}+\frac{\Pi}{4}\Big( Q_{\nu\alpha\beta}u_{\mu}+Q_{\nu\mu\alpha}u^{\alpha} \Big)
\eeq

In particular let us mention the absence of the velocity related (kinematic) shear $\tilde{\sigma}_{\mu\nu}$.

Variation of the action (\ref{actionShear}) with respect to the rest of the fluid variables, yields the system of equations 
\bs
\ba
0 & = & \delta I/\delta J^\mu = \mu u_\mu + \varphi_{,\mu} + s\theta_{,\mu} + \beta_A\alpha^A{}_{,\mu}-\frac{1}{4n}\mathcal{Z}_{\alpha} h^{\alpha}{}_{\mu} \,,  \\
0 & = & \delta I/\delta \varphi = - \mathring{\nabla}_\mu J^\mu\,,  \\
0 & = & \delta I/\delta \theta  = - \mathring{\nabla}_\mu\lp s J^\mu\rp\,, \\ 
0 & = & \delta I/\delta s  = -\sqrt{-g}\frac{\partial\rho}{\partial s} + \theta_{,\mu} J^\mu\,, \\
0 & = & \delta I/\delta \alpha^A = -\partial_{\mu}(J^{\mu}\beta_{A})\,, \\
0 & = & \delta I/\delta \beta_A = J^\mu\alpha^A{}_{,\mu}\,,  \\
0 & = & \delta I/\delta \Sigma = -\frac{\partial\rho}{\partial \Sigma} -\frac{1}{4} Q_{\alpha\mu\nu}u^{\alpha}u^{\mu}u^{\nu}  -\frac{1}{4 d} Q_{\mu}u^{\mu} \,,  \\
0 & = & \delta I/\delta \Pi = -\frac{\partial\rho}{\partial \Pi}  -\frac{1}{4} Q_{\alpha\mu\nu}u^{\alpha}u^{\mu} u^{\nu}   -\frac{1}{4 } q_{\mu}u^{\mu}\,.  
\ea
\es
where
\beq
\mathcal{Z}_{\alpha}=(\Sigma+\Pi)\Big( Q_{\alpha\beta\gamma}u^{\beta}u^{\gamma}+2 Q_{\beta\alpha\gamma}u^{\beta}u^{\gamma}\Big) +\frac{\Sigma}{d}Q_{\alpha}+\Pi q_{\alpha} \label{ZZ}
\eeq

In particular, the last two equations of the above system when subtracted give
\beq
\frac{\partial\rho}{\partial \Sigma} -
\frac{\partial\rho}{\partial \Pi} =\frac{1}{4}\Big( q_{\mu}-\frac{Q_{\mu}}{d}\Big) u^{\mu}
\eeq
Let us observe that  that for Weyl non-metricity, i.e. $Q_{\alpha\mu\nu}=\frac{1}{d}Q_{\alpha}g_{\mu\nu}$ it follows that $\frac{\partial\rho}{\partial \Sigma}=\frac{\partial\rho}{\partial \Pi}=0$  verifying  the fact that the Weyl part (i.e. trace) of non-metricity is related to the dilation part, as we showed earlier and not to that of shear.

\section{General Case with full Hypermomentum}
\label{general}

It is now trivial to generalize the previous considerations to the general case where all parts of hypermomentum of the fluid are included, namely all dilation, spin and shear parts are non-vanishing.  The full action is additive and reads
\begin{gather}
I = \int d^d x \Big[ -\sqrt{-g}\rho(n,s,h,\sigma, \zeta,\Sigma,\Pi) + {J}^\mu\Big( \varphi_{,\mu} + s\theta_{,\mu} + \beta_A \alpha^A{}_{,\mu}\Big)\nonumber \\ -\frac{1}{4}Q_{\mu}\mathcal{D}^{\mu}
 -\frac{1}{2}\Big( -2 S_{\mu}+\frac{1}{2}q_{\mu}-\frac{1}{2}Q_{\mu}\Big) \sigma^{\mu}+\frac{1}{2}t_{\mu}\zeta^{\mu}  
-\frac{1}{4}(\Sigma^{\alpha}+\Pi^{\alpha})Q_{\alpha\mu\nu}u^{\mu}u^{\nu}-\frac{1}{4d}\Sigma^{\mu}Q_{\mu}-\frac{1}{4}\Pi^{\mu}q_{\mu}  \Big] \label{actionFull}
\end{gather}
with the first line containing the usual  terms for relativistic perfect fluids \cite{Brown:1992kc} while the second contains all pieces the excite the hypermomentum (making it a hyperfluid). Variation with respect to the connection reproduces the full hypermomentum (\ref{FullHyper}), while variation with respect to the metric gives us the energy-momentum tensor
    \begin{gather}
        T_{\mu\nu}=\left[ \rho+\dot{f}+f \tilde{\Theta}-\frac{1}{2}(\Sigma +\Pi)Q_{\alpha\beta\gamma}u^{\alpha}u^{\beta}u^{\gamma}-\frac{1}{2 \sqrt{-g}}\zeta^{\alpha}t_{\alpha} \right] u_{\mu}u_{\nu}+\left[   p+ \dot{F}+F \tilde{\Theta}-\frac{(\Sigma +\Pi)}{2(d-1)}\tilde{\Theta}+\frac{1}{2 \sqrt{-g}}\zeta^{\alpha}t_{\alpha}   \right]h_{\mu\nu} \nonumber \\
       -\frac{(\sigma+\Pi)}{2} \tilde{\sigma}_{\mu\nu} +\frac{(\sigma+\Pi)}{2}N_{\alpha(\mu\nu)}u^{\alpha}+\frac{1}{2}(\sigma-2 \Sigma -\Pi)u_{(\mu}\tilde{a}_{\nu)}+\frac{(\sigma+\Pi)}{2} u_{(\mu}A_{\nu)}
        -\frac{1}{2} u_{(\mu}\partial_{\nu)}(\sigma+\Pi)
        \nonumber \\
        +(\Sigma+\Pi)u^{\alpha}u^{\beta}(N_{\alpha(\mu|\beta|}u_{\nu)}-Q_{\alpha\beta(\mu}u_{\nu)})-2 \zeta u^{\rho}\epsilon_{\rho\alpha\beta(\mu}S_{\nu)}{}^{\alpha\beta}
    \end{gather}
which of course has the imperfect fluid form. In the above we have abbreviated 
\beq
f=\frac{1}{2}\left( h-\sigma -\frac{(d-1)}{d}\Sigma -\Pi \right) ,\quad  F=-\frac{1}{2}\left( h-\sigma +\frac{1}{d}\Sigma \right)
\eeq
and have defined the generalized hyper-pressure
\beq
p=n \frac{\partial \rho}{\partial n}+h \frac{\partial \rho}{\partial h}+\sigma \frac{\partial \rho}{\partial \sigma}+\zeta \frac{\partial \rho}{\partial \zeta}+\Sigma \frac{\partial \rho}{\partial \Sigma}+\Pi \frac{\partial \rho}{\partial \Pi}-\rho
\eeq
The viscous density and pressure are computed as
\begin{gather}
    \hat{\rho}=\rho +\frac{1}{2}\left[ u^{\alpha}\partial_{\alpha}h-\frac{(d-1)}{d} u^{\alpha}\partial_{\alpha}\Sigma \right]+\frac{1}{2}\left[ h-\sigma -\frac{(d-1)}{d}\Sigma -\Pi \right]\tilde{\Theta}+\frac{(\sigma +\Pi)}{2}\left[ \frac{1}{2}Q_{\alpha\beta\gamma}u^{\alpha}u^{\beta}u^{\gamma}-A_{\mu}u^{\mu} \right]-\frac{1}{2 \sqrt{-g}}\zeta^{\alpha}t_{\alpha}
\end{gather}
and
\begin{gather}
    (d-1)\hat{p}=(d-1)p +\frac{(d-1)}{2}\left[ -u^{\alpha}\partial_{\alpha}h+ u^{\alpha}\partial_{\alpha}\sigma-\frac{1}{d}u^{\mu}\partial_{\mu}\Sigma \right]+\frac{1}{2}\left[ -(d-1)h+(d-2)\sigma -\frac{(d-1)}{d}\Sigma -\Pi \right]\tilde{\Theta} \nonumber \\+\frac{(\sigma +\Pi)}{2}\left[ \frac{1}{2}Q_{\alpha\beta\gamma}u^{\alpha}u^{\beta}u^{\gamma}-A_{\mu}u^{\mu} \right]-\frac{1}{2 \sqrt{-g}}\zeta^{\alpha}t_{\alpha}
\end{gather}

The associated heat flux is
\begin{gather}
\mathcal{Q}_{\lambda}=\frac{1}{4}(\sigma-2 \Sigma-\Pi)\tilde{a}_{\lambda}+\frac{(\sigma+\Pi)}{4} A_{\lambda}-\frac{1}{4}\partial_{\lambda}(\sigma +\Pi)+\frac{(\sigma+\Pi)}{4}\Big( A_{\mu}u^{\mu}-Q_{\alpha\beta\gamma}u^{\alpha}u^{\beta}u^{\gamma}\Big)u_{\lambda}-\frac{1}{4}u_{\lambda}u^{\mu}\partial_{\mu}(\sigma+\Pi) \nonumber \\
-\frac{\Pi}{2}u^{\alpha}u^{\beta}N_{\alpha(\beta\lambda)}+\frac{(\Sigma+\Pi)}{2}(N_{\alpha\lambda\beta}u^{\alpha}u^{\beta}-Q_{\alpha\beta\lambda}u^{\alpha}u^{\beta})-\frac{(\Sigma+\Pi)}{4}Q_{\alpha\beta\gamma}u^{\alpha}u^{\beta}u^{\gamma}u_{\lambda}\nonumber \\-\frac{\sigma}{4}(Q_{\lambda\alpha\beta}+2 S_{\lambda\alpha\beta})u^{\alpha}u^{\beta}+\zeta u^{\rho}\epsilon_{\rho\alpha\beta\lambda}S_{\mu}{}^{\alpha\beta}u^{\mu}
\end{gather}
and the anisotropic stresses are found to be
\begin{gather}
\pi_{\alpha\beta}=-\frac{(\sigma+\Pi)}{2}\tilde{\sigma}_{\alpha\beta}+\frac{(\sigma+\Pi)}{2}\left[ u^{\lambda}N_{\lambda(\alpha\beta)}+u^{\mu}u^{\lambda}(N_{\lambda(\alpha|\mu|}u_{\beta)}+N_{\lambda\mu(\alpha}u_{\beta)})+\frac{1}{2}(Q_{\lambda\mu\nu}u^{\lambda}u^{\mu}u^{\nu})u_{\alpha}u_{\beta} \right] \nonumber \\
+\frac{h_{\alpha\beta}}{(d-1)}\frac{(\sigma+\Pi)}{2 }\Big[ A_{\mu}u^{\mu}-\frac{1}{2}Q_{\lambda\mu\nu}u^{\lambda}u^{\mu}u^{\nu} \Big]+\frac{h_{\alpha\beta}}{(d-1)} \frac{2}{ \sqrt{-g}}\zeta^{\alpha}t_{\alpha}-2 \zeta u^{\rho}\epsilon_{\rho\mu\nu(\alpha}\Big( S_{\beta)}{}^{\mu\nu}+u_{\beta)}S_{\lambda}{}^{\mu\nu}u^{\lambda} \Big) \label{pfu}
\end{gather}
It is worthwhile to observe that in the expression for the stresses only one of the shear parts (namely $\Pi$) contributes and furthermore the spin-shear $(\sigma+\Pi)$ combination appears throughout (\ref{pfu}).

Variation of the action (\ref{actionFull}) with respect to the rest of the fluid variables, yields the system of equations 
\bs
\ba
0 & = & \delta I/\delta J^\mu = \mu u_\mu + \varphi_{,\mu} + s\theta_{,\mu} + \beta_A\alpha^A{}_{,\mu}-\frac{1}{4n}\tau_{\alpha}h^{\alpha}{}_{\mu} \,,  \\
0 & = & \delta I/\delta \varphi = - \mathring{\nabla}_\mu J^\mu\,,  \\
0 & = & \delta I/\delta \theta  = - \mathring{\nabla}_\mu\lp s J^\mu\rp\,, \\ 
0 & = & \delta I/\delta s  = -\sqrt{-g}\frac{\partial\rho}{\partial s} + \theta_{,\mu} J^\mu\,, \\
0 & = & \delta I/\delta \alpha^A = -\partial_{\mu}(J^{\mu}\beta_{A})\,, \\
0 & = & \delta I/\delta \beta_A = J^\mu\alpha^A{}_{,\mu}\,,  
\\
0 & = & \delta I/\delta h =-\frac{\partial\rho}{\partial h} -\frac{1}{4} Q_{\mu}u^{\mu} \,,   \label{hh}
\\
0 & = & \delta I/\delta \sigma = -\frac{\partial\rho}{\partial \sigma} -\frac{1}{2} \Big( -2 S_{\mu}-\frac{1}{2}Q_{\mu}+\frac{1}{2}q_{\mu}\Big)u^{\mu} \,,
\\
0 & = & \delta I/\delta \zeta=-\frac{\partial\rho}{\partial \zeta} +\frac{1}{2} t_{\mu}u^{\mu} \,,  
\\
0 & = & \delta I/\delta \Sigma = -\frac{\partial\rho}{\partial \Sigma} -\frac{1}{4} Q_{\alpha\mu\nu}u^{\alpha}u^{\mu}u^{\nu}  -\frac{1}{4 d} Q_{\mu}u^{\mu} \,,  \\
0 & = & \delta I/\delta \Pi = -\frac{\partial\rho}{\partial \Pi}  -\frac{1}{4} Q_{\alpha\mu\nu}u^{\alpha}u^{\mu} u^{\nu}   -\frac{1}{4 } q_{\mu}u^{\mu}\,.   \label{QQ}
\ea
\es
where 
\beq
\tau_{\alpha}=h Q_{\alpha}+4\xi_{\alpha}+2 \zeta t_{\alpha}+\mathcal{Z}_{\alpha}
\eeq
with $\xi_{\alpha}$ and $\mathcal{Z}_{\alpha}$ being given by (\ref{XX}) and (\ref{ZZ}) respectively.

It should be noted that all scalar quantities formed from the torsion and non-metricity vectors and $u^{\mu}$ are expressible  in terms of partial derivatives of the density $\rho$ of the fluid,  as a consequence of the above equations of motion. Indeed, equations (\ref{hh})-(\ref{QQ}) form a simple system of $5$ equations which we may trivially solve in terms of the partial derivatives of $\rho$, yielding finally

\bs
\ba
 Q_{\mu}u^{\mu} & = &-4 \frac{\partial \rho}{\partial h}  \,,  \\
q_{\mu}u^{\mu} & = & 4 \left( -\frac{\partial \rho}{\partial \Sigma}+\frac{\partial \rho}{\partial \Pi}-\frac{1}{d}\frac{\partial \rho}{\partial h} \right)  \,, \\
S_{\mu}u^{\mu} & = &  \frac{\partial \rho}{\partial \sigma}+\frac{(d-1)}{d}\frac{\partial \rho}{\partial h}-\frac{\partial \rho}{\partial \Sigma}  +\frac{\partial \rho}{\partial \Pi}  \, , 
\\
t_{\mu}u^{\mu} & = & 2 \frac{\partial \rho}{\partial \zeta}   
  \,.
\ea
\es
as a bonus we also find the totally symmetric non-metricity piece:
\beq
Q_{\alpha\beta\gamma}u^{\alpha}u^{\beta}u^{\gamma}  =4 \left(  \frac{1}{d}\frac{\partial \rho}{\partial h}- \frac{\partial \rho}{\partial \Sigma} \right)
\eeq
Using these results it is then straightforward to see that the on-shell action reads
\beq
I_{on-shell}=\int d^{d}x \sqrt{-g}p
\eeq
namely, it is determined, once again, only by the (hyper)-pressure of the fluid\footnote{Quite interestingly the on-shell action continues to be given by the (hyper)-pressure also to the formulation other kinds of hyperfluids \cite{Obukhov:2023yti}.}. Of course as we have already pointed action the value of the on-shell action is not unique and can be changed upon adding surface terms to (\ref{actionFull}) (see \cite{Brown:1992kc}).

Furthermore, using the above relations we can also evaluate the expression of the quantity $A_{\mu}u^{\mu}$, which appears in many places, it reads
\beq
A_{\mu}u^{\mu}=2 \left( \frac{1}{d}\frac{\partial \rho}{\partial h}+\frac{\partial \rho}{\partial \sigma} + \frac{\partial \rho}{\partial \Sigma}-\frac{\partial \rho}{\partial \Pi}   \right) 
\eeq

Let us emphasize once more the fact that the viscous fluid characteristics, namely the heat flux, stresses and viscous density and pressure we obtain are of the same form with the ones coming from the $1st$-order development. However, is is quite remarkable that in our case all coefficient functions are determined and have an origin; the hypermomentum. We find this result astonishing. A further investigation on this connection between  viscous fluids and hypermomentum would be certainly worth undertaking.

\section{On the Perfect Hyperfluid sector}
\label{perfect}

As we have shown, in general hyperfluids are imperfect in the sense that the associated energy-momentum tensor does not have the simple perfect fluid form but rather that of an imperfect one. It is then natural to ask whether it is possible to construct a hyperfluid model where the energy-momentum tensor retains its perfect fluid form. The answer is affirmative as we show below. The first thing to observe before developing this possibility is the fact that shear and bulk viscosities as well as anisotropic stresses and heat fluxes   originate from the metric variations of  terms that involve contractions of densities with $q_{\mu}$, $Q_{(\alpha\mu\nu)}$ and $t_{\mu}$. On the other hand, terms like $\mathcal{X}^{\mu}Q_{\mu}$ and $\mathcal{Y}^{\mu}S_{\mu}$ do not reproduce such 'imperfect' terms. In particular, the former  contributes a part that has a perfect fluid form whereas the latter does not bring any explicit modification to the form of the energy-momentum tensor at all.
\footnote{This is so because both $S_{\mu}$ and $\mathcal{Y}^{\mu}$ are metric independent. Consequently  $\mathcal{Y}^{\mu}S_{\mu}$ is also metric independent and as such does not contribute to the energy-momentum tensor.} It thus becomes clear that adding these two terms into our fluid action would then reproduce a perfect fluid form for the energy-momentum tensor. With the above observations we are now able to write down the action for the perfect hyperfluid:
\begin{gather}
I_{perfect} = \int d^d x \Big[ -\sqrt{-g}\rho(n,s,X,Y) + {J}^\mu\Big( \varphi_{,\mu} + s\theta_{,\mu} + \beta_A \alpha^A{}_{,\mu}\Big) -\frac{1}{4}Q_{\mu}\mathcal{X}^{\mu}+S_{\mu}\mathcal{Y}^{\mu}\Big] \label{rwq}
\end{gather}
where we introduced the densities $\mathcal{X}^{\mu}:=\sqrt{-g}X u^{\mu}=\sqrt{-g}X^{\mu}$ and $\mathcal{Y}^{\mu}:=\sqrt{-g}Y u^{\mu}=\sqrt{-g}Y^{\mu}$. The associated hypermomentum to (\ref{rwq}), reads
\beq
\Delta_{\lambda}{}^{\mu\nu}=\delta_{\lambda}^{\mu}(X^{\nu}+Y^{\nu})-\delta_{\lambda}^{\nu}Y^{\mu}
\eeq
Let us note that this is a hybrid kind of hypermomentum containing pieces from all dilation, spin and shear parts. Indeed, it immediately follows that
\beq
\Delta_{\nu}= X_{\nu}+\frac{(d-1)}{d}Y_{\nu}
\eeq
\beq
\Delta_{[\alpha\mu]\nu}=-g_{\nu[\alpha}Y_{\mu]}
\eeq
\beq
\hat{\Delta}_{\alpha\mu\nu}=\frac{1}{d}g_{\alpha\mu}Y_{\nu}-g_{\nu(\alpha}Y_{\mu)}
\eeq
It is also worth mentioning that three pieces could be excited  only by including $S_{\mu}\mathcal{Y}^{\mu}$ alone. Then variation of (\ref{rwq}) with respect to the metric gives the energy-momentum tensor, which has the perfect fluid form as stated:
\beq
T^{\mu\nu}=\rho_{m}u^{\mu}u^{\nu}+p_{m}h^{\mu\nu} \label{Tpfa}
\eeq
In the above we have set
\beq
\rho_{m}=\rho+\frac{1}{2 \sqrt{-g}}\partial_{\alpha}(\sqrt{-g}X u^{\alpha})
\eeq
\beq
p_{m}=p-\frac{1}{2 \sqrt{-g}}\partial_{\alpha}(\sqrt{-g}X u^{\alpha}) 
\eeq
where 
\beq
p:=n\frac{\partial \rho}{\partial n}+X\frac{\partial \rho}{\partial X}+Y\frac{\partial \rho}{\partial Y}- \rho  \label{Kp}
\eeq
Let us observe that in our expression (\ref{Tpfa}) there appears no explicit dependence on the hypermomentum density $\mathcal{Y}^{\mu}$, however there is an implicit dependence through the density $\rho(n,s,X,Y)$ as well as the hyperpressure (\ref{Kp}). For completeness let us also report the rest of the fluid's equations of motion:
\bs
\ba
0 & = & \delta I/\delta J^\mu = \mu u_\mu + \varphi_{,\mu} + s\theta_{,\mu} + \beta_A\alpha^A{}_{,\mu}-\frac{1}{n}\Big( -\frac{X}{4}Q_{\alpha}+Y S_{\alpha}\Big) h^{\alpha}{}_{\mu} \,,  \\
0 & = & \delta I/\delta \varphi = - \mathring{\nabla}_\mu J^\mu\,,  \\
0 & = & \delta I/\delta \theta  = - \mathring{\nabla}_\mu\lp s J^\mu\rp\,, \\ 
0 & = & \delta I/\delta s  = -\sqrt{-g}\frac{\partial\rho}{\partial s} + \theta_{,\mu} J^\mu\,, \\
0 & = & \delta I/\delta \alpha^A = -\partial_{\mu}(J^{\mu}\beta_{A})\,, \\
0 & = & \delta I/\delta \beta_A = J^\mu\alpha^A{}_{,\mu}\,,  \\
0 & = & \delta I/\delta X = -\sqrt{-g}\frac{\partial\rho}{\partial X} -\frac{1}{4}Q_{\mu}u^{\mu}\sqrt{-g}  \,,  \\
0 & = & \delta I/\delta Y = -\sqrt{-g}\frac{\partial\rho}{\partial Y}+S_{\mu}u^{\mu}\sqrt{-g}  \,.  
\ea
\es

Having discussed also the  Perfect Hyperfluid sector we can now focus on some applications of our formulation.

\section{Applications}
\label{cosmo}

As an application let us consider the the Theory
\beq
S=\frac{1}{2 \kappa}\int d^{n}x \sqrt{-g}R +S_{spin}
\eeq
where $R$ is the scalar curvature of the full connection and $S_{spin}$  the matter part of a pure spin hyperfluid as given by (\ref{actionSpin}). Let us note that because of the projective invariance of $R$ a dilation fluid is not possible in this case. In order to be able to allow for a dilation part, quadratic terms in torsion and non-metricity must be added to break the invariance.
Now back to our model, varying with respect to the metric we get
\beq
R_{(\mu\nu)}-\frac{R}{2}g_{\mu\nu}=\kappa T_{\mu\nu} \label{mefeq}
\eeq
which resemble Einstein's field equations but remember that now the Ricci tensor and scalar have additional parts due to torsion and non-metricity. Varying with respect to the
connection we obtain 
\beq
P_{\lambda}{}^{\mu\nu}=\kappa \Delta_{\lambda}{}^{\mu\nu} \label{Pal}
\eeq
where
\beq
	P_{\lambda}{}^{\mu\nu}=-\frac{\nabla_{\lambda}(\sqrt{-g}g^{\mu\nu})}{\sqrt{-g}}+\frac{\nabla_{\sigma}(\sqrt{-g}g^{\mu\sigma})\delta^{\nu}_{\lambda}}{\sqrt{-g}} \\
	+2(S_{\lambda}g^{\mu\nu}-S^{\mu}\delta_{\lambda}^{\nu}+g^{\mu\sigma}S_{\sigma\lambda}{}{}^{\nu}) \label{Cone}
	\eeq
and the hypermomentum on the right-hand side is that of a pure spin hyperfluid, namely it is given by (\ref{spinD}).

Taking the contractions of the above field equations and exploiting the projective symmetry to set $q_{\mu}=0$ we find
\beq
S_{\mu}=\frac{(n-1)}{4(n-2)\sqrt{-g}}\kappa\sigma_{\mu} ,\quad Q_{\mu}=0
\eeq
Then the distortion tensor is found to be (see \cite{iosifidis2019exactly})
\beq
N_{\alpha\mu\nu}=-\frac{\kappa}{ (n-2)\sqrt{-g}}\sigma_{[\alpha}g_{\mu]\nu}-\frac{\kappa}{2 \sqrt{-g}}\epsilon_{\mu\nu\alpha\beta}\zeta^{\beta}
\eeq
from which we immediately extract the form of torsion and non-metricity as
\beq
S_{\mu\nu\alpha}=\frac{\kappa}{2 (n-2)\sqrt{-g}}\sigma_{[\mu}g_{\nu]\alpha}-\frac{\kappa}{2 \sqrt{-g}}\epsilon_{\mu\nu\alpha\beta}\zeta^{\beta}
\eeq
and 
\beq
Q_{\alpha\mu\nu}=0
\eeq
we see that the full non-metricity vanishes. Therefore, in this minimalistic model spin excites only torsion. To torsion pseudovector is then easily found to be
\beq
t_{\mu}=-\frac{3}{\sqrt{-g}}\kappa\zeta_{\mu}
\eeq
Now, from equations (\ref{Sss1}) and (\ref{Sss2}) it follows that
\beq
\frac{ \partial \rho}{\partial \sigma}=-\frac{(d-1)}{4(d-2)}\kappa\sigma , \qquad \frac{ \partial \rho}{\partial \zeta}=\frac{3}{2} \kappa\zeta
\eeq
which, when combined, imply that the density is of the form
\beq
\rho(n, \sigma, \zeta)=-\frac{(d-1)}{8(d-2)}\kappa\sigma^{2}+\frac{3}{4}\kappa \zeta^{2}+f(n) 
\eeq
where $f(n)$ depends only on the particle number density. This function is then naturally associated to the perfect fluid component of the hyperfluid and has the power-law form $f(n)\propto n^{a}$ for barotropic perfect fluids. Note that from the latter for one then obtains the equation of state for the perfect fluid pressure. Let us remark that for the spin components of the hyperfluid, in this case, the 'spin equations of state' are of the 'stiff' matter type, namely $p_{x}=\rho_{x}$, for $x=\sigma, \zeta$.
The hyperpressure is then given by,
\beq
p=n \frac{ \partial \rho}{\partial n}-f(n) -\frac{(d-1)}{8(d-2)}\kappa\sigma^{2}+\frac{3}{4}\kappa \zeta^{2}
\eeq
and this can have a barotropic form $p=w \rho$ only if the perfect fluid component is that of stiff matter. We may set $\rho_{pf}=f(n)$ and then
\beq
p_{pf}=n \frac{ d f}{d n}-f(n)=n \frac{ d \rho_{pf}}{d n}-\rho_{pf}
\eeq
denotes the perfect fluid contribution to the pressure.

Accordingly, the viscous density and pressure  are found to be
\beq
\hat{\rho}=\rho_{pf}+\frac{(d-1)}{8(d-2)}\kappa\sigma^{2}-\frac{3}{2}\kappa \zeta^{2}-\kappa\frac{\sigma}{2}\tilde{\Theta} \label{frho}
\eeq
and
\beq
(d-1)\hat{p}=(d-1)p_{pf}+(d-3)\left[ -\frac{(d-1)}{8(d-2)}\kappa\sigma^{2}+\frac{3}{2}\kappa \zeta^{2} \right]+\frac{(d-1)}{2} \dot{\sigma}+\frac{\sigma}{2}(d-2)\tilde{\Theta} \label{fpp}
\eeq
Additionally, the heat flux and anisotropic stresses for this case, read
\beq 
\mathcal{Q}_{\lambda}=\frac{1}{4}\Big( \sigma \tilde{a}_{\lambda}-\partial_{\lambda}\sigma -\dot{\sigma}u_{\lambda} \Big) \label{flux}
\eeq
and
\beq \label{astress}
\pi_{\alpha\beta}=-\frac{\sigma}{2}\tilde{\sigma}_{\alpha\beta}
\eeq
It is quite remarkable that all extra terms have cancelled out and in the end the forms of the heat flux and stresses are exactly those we have for viscous fluids! Note that $\zeta$ only contributes to the viscous density and pressure and not to heat flux and stresses.

One could then, in principle, relate the microscopic hypermomentum variables to the temperature and also obtain the values of thermal conductivity, coefficients of shear and bulk viscosity and so on. For instance recall that for $1^{st}$-order dissipative viscous fluids we have the heat flux given by\footnote{T is the temperature and $\mathcal{K}(T)$ is the thermal conductivity.}
\beq
\mathcal{Q}_{\lambda}=-\mathcal{K}\Big( \tilde{a}_{\lambda}+\partial_{\lambda}T +\dot{T}u_{\lambda} \Big) \label{Qk}
\eeq
Comparing then with $(\ref{flux})$ one easily finds that $\sigma  =C_{0} e^{-T} $, $C_{0}=const.$ and $\mathcal{K}(T)=-\frac{C_{0}}{4} e^{-T}$. Given that $\mathcal{K}(T)>0$ we conclude that $C_{0}$ must be negative and so $\sigma <0$. This, in turn, implies that the viscosity coefficient $\zeta_{visc.}=-\sigma/2$ appearing in (\ref{astress}) is positive as it should.  Of course, in order to have a complete picture and determine every coefficient one needs to have the full hyperfluid (i.e. with the dilation and shear currents as well).

\subsection{Pure Spin Fluid Cosmology}
Let us now briefly touch upon the cosmological implications of such a hypermomentum induced viscous fluid\footnote{For recent study on the Cosmology of general first-order viscous fluids see \cite{Bemfica:2022dnk}. }. We will therefore fix $d=4$. Considering a homogeneous,  isotropic and flat  FLRW Universe, the covariant forms of allowed torsion and non-metricity read \cite{Iosifidis:2020gth}
\beq
S_{\mu\nu\alpha}^{(n)}=2u_{[\mu}h_{\nu]\alpha}\Phi(t)+\epsilon_{\mu\nu\alpha\rho}u^{\rho}P(t)\delta_{n,4} \label{isotor}
\eeq
and 
	\beq
	Q_{\alpha \mu \nu}  = A(t) u_\alpha h_{\mu \nu} + B(t) h_{\alpha(\mu} u_{\nu)} + C(t) u_\alpha u_\mu u_\nu
	\eeq
 respectively. In the above, the functions  $\Phi, P$ describe the torsion and 
	 $A,B$ and $C$ the non-metricity. In such a highly constrained geometry, all heat flux, stresses, viscosity etc. identically vanish and both the canonical and metric energy-momentum tensors have necessarily the perfect fluid form, viz.
  \beq
t_{\mu\nu}=\rho u_{\mu}u_{\nu}+p h_{\mu\nu} \label{tt}
  \eeq
and 
\beq
T_{\mu\nu}=\hat{\rho}u_{\mu}u_{\nu}+\hat{p} h_{\mu\nu} \label{TT}
\eeq
It is then worth mentioning that the density and pressure appearing in the above energy-momentum tensor are in fact the 'viscous corrected' density and pressure respectively and not the 'perfect ones'. The generic (i.e. non FLRW oriented) Lagrangian formulation we developed here reveals this distinction, something that would be otherwise hidden,  had we started right away with the isotropic ansatz. Plugging, (\ref{tt}) and (\ref{TT}), along with the hypermomentum form,  in the conservation laws (\ref{cc1}) and (\ref{cc2}) we find
\beq
	\dot{\rho}+3H(\rho+p)	=\frac{\sigma}{2}( R_{\mu\nu}- \breve{R}_{\mu\nu})u^{\mu}u^{\nu}+\frac{1}{2}(\rho-\hat{\rho})(A+C) \label{r2}
	\eeq
		\beq
	 \sigma \frac{B}{2}=2 (p_{c}-p) 	\label{r}
	\eeq
	\beq
3 \sigma \frac{B}{2}=2(\rho-\hat{\rho}) \label{p}
	\eeq
	\beq
	(\rho-\hat{\rho}) =3(p-\hat{p}) 
	\eeq
Now given that in our case non-metricity vanishes, setting $0=A=B=C$, to the the above results in
\beq
\hat{p}=p, \quad \hat{\rho}=\rho \label{hphh}
\eeq
namely that the hyperfluid is hypermomentum preserving. We note the perfect agreement of our result with \cite{Iosifidis:2023but} where it was concluded that a pure spin hypermomentum preserving hyperfluid is in general not possible and only allowed  if non-metricity vanishes. Our example here falls exactly in this peculiar case.

Furthermore, since $\breve{R}_{\mu\nu}=-R_{\mu\nu}$ for zero non-metricity, the modified continuity equation (\ref{r2}) simplifies to
\beq
\dot{\rho}+3H(\rho+p)	=\frac{\kappa}{2}\sigma (\rho+3 p) \label{Prr}
\eeq
where we have also used $R_{\mu\nu}u^{\mu}u^{\nu}=\frac{\kappa}{2}(\rho+3 p)$ as can be easily proved by contracted the metric field equations with $u^{\mu}u^{\nu}$. Equation (\ref{Prr}) is also in perfect agreement with the one derived in \cite{kranas2019friedmann}. However, with our Lagrangian formulation we have gained a very illuminating view, namely we have found out that the pressure the density $\rho$ and pressure $p$ that appear in (\ref{Prr}) are not quite the pure perfect fluid components but rather these are the viscous density and pressure respectively that contain in them hypermomentum contributions. Indeed, combing (\ref{hphh}) with (\ref{frho}) and (\ref{fpp}) it follows that 
\beq
\rho=\rho_{pf}+3\left( \frac{\kappa\sigma^{2}}{16}-\frac{1}{2}H  \sigma \right)  \label{rho11}
\eeq
\beq
p=p_{pf}-\left( \frac{\kappa\sigma^{2}}{16}-\frac{1}{2}H \sigma \right)+\frac{\dot{\sigma}}{2} \label{p22}
\eeq
where we have considered only the parity even sector, namely we have set $\zeta=0$ for simplicity. It cannot be too strongly emphasized the fact that the viscous modifications that appear in the above equations appeared dynamically (i.e. without adding them by hand) and have an origin namely they come from the microproperties of the fluid as they are encoded in the hypermomentum tensor. Going one step further, on substituting the above forms of density and pressure in the conservation law (\ref{Prr}) we get the evolution equation:
\beq
\Big[\dot{\rho}_{pf}+3H (\rho_{pf}+ p_{pf})-\frac{\kappa \sigma}{2}(\rho_{pf}+3 p_{pf}) \Big] -\frac{3 \sigma}{8} \Big[ \kappa \dot{\sigma}+4 \dot{H}-\kappa\sigma H +8 H^{2}\Big] =0\label{sigtor}
\eeq
The collection of the above results combined with the modified  Friedmann equations that can be extracted from the metric field equations (\ref{mefeq}), describe then a torsional cosmology. For completeness let us also include these modified Friedmann equations. They read
\beq
H^{2}=\frac{\kappa \rho}{3}+H\kappa \frac{\sigma}{2} -\frac{\kappa^{2}\sigma^{2}}{16}
\eeq
and
\begin{gather}
    \frac{\ddot{a}}{a}=-\frac{\kappa}{6}(\rho + 3 p)+\frac{\kappa}{4}(\dot{\sigma}+H \sigma)
\end{gather}
We may now also use relations (\ref{rho11}) and (\ref{p22}), and bring the Friedmann equations to the form
\beq
H^{2}=\frac{\kappa \rho_{pf}}{3} \label{fq1}
\eeq
\begin{gather}
    \frac{\ddot{a}}{a}=-\frac{\kappa}{6}(\rho_{pf} + 3 p_{pf})+\frac{\kappa}{4}H \sigma \label{fq2}
\end{gather}
It is quite remarkable that all hypermomentum variables have canceled out  in  the first Friedmann equation and in the end it retained its usual form. Furthermore we see that the acceleration equation have received a viscous term \cite{maartens1995dissipative,brevik2017review,brevik2017viscous,medina2019viscous} that is exactly the one appearing in viscous cosmologies! Note that the term $\kappa \sigma $ has a clear interpretation as the bulk viscosity coefficient (up to numerical factors)
Furthermore, differentiating (\ref{fq1}) with respect to $t$ and on using also (\ref{fq2}), plugging all back at (\ref{sigtor}) we find the evolution equation for $\sigma$\footnote{This is obtained in the premise that $\sigma \neq 0$. One also has the trivial solution $\sigma=0$ as a possibility, which however is of no interest to us.}:
\beq
 \dot{\sigma}=-\frac{2}{3}(\rho_{pf}+3p_{pf}) \label{sevol}
\eeq
We also get the continuity equation for the perfect fluid component
\beq
\dot{\rho}_{pf}+3H (\rho_{pf}+ p_{pf})=\frac{\kappa}{2}\sigma \rho_{pf} \label{ccom}
\eeq

Let us now find a solution to the above system. Consider the case of a conventional perfect fluid component with the usual barotropic equation of state  $p_{pf}=w \rho_{pf}$ and let us make the ansatz $\kappa\sigma=\lambda H$ where $\lambda$ is some dimensionless constant to be evaluated by imposing the consistency conditions. Using the first Friedmann equation and the the above ansatz combined with (\ref{sevol}) we arrive at
\beq
\dot{H}+\frac{2(1+3w)}{\lambda}H^{2}=0 \label{HH1}
\eeq
Then under these conditions the continuity equation (\ref{ccom}) becomes
\beq
\dot{H}+\Big[\frac{3}{2}(1+w)-\frac{\lambda}{4}\Big]H^{2}=0 \label{HHH}
\eeq
So, for the last two equations to be consistent with one another, it must hold that 
\beq
\lambda^{2}-6  (1+w)\lambda+8(1+3 w)=0
\eeq
with determinant $\Delta=4(1-3 w)^{2} \geq 0$ and solutions
\beq
\lambda_{1,2}=3(1+w)\pm |1-3 w| \label{quadsolut}
\eeq
Let 
\beq
\gamma=\frac{2(1+3w)}{\lambda_{1,2}}
\eeq
Then integrating (\ref{HH1}) we get
\beq
H=\frac{1}{\gamma t +c_{1}}
\eeq
and integrating once more we get the nice powerlaw solutions for the scale factor
\beq
a(t)=c_{2}(\gamma t+c_{1})^{\frac{1}{\gamma}}
\eeq
where $c_{1,2}$ are integration constants. The value of the exponent depends solely on the barotropic index $w$. From the last two equations above we can also infer the effective barotropic index $w_{eff}$ for the net contribution. We easily find\footnote{Using the fact that $w_{eff}$ is defined through $H^{2}\propto \frac{1}{a^{3(1+w_{eff})}}$.}
\beq
w_{eff}=\frac{2}{3}\gamma -1
\eeq
For $w \neq 1/3$ we find 
 the two solutions are $\lambda_{1}=4$ and $\lambda_{2}=2(1+3w)$, for both signs of (1-3w) and the corresponding barotropic indices read
\beq
w_{eff}^{(1)}= w-\frac{2}{3}\quad, \quad w_{eff}^{(2)}=-1/3 
\eeq
    and for the degenerate case of radiation (i.e. $w=1/3$) the effective index reads
    \beq
w_{eff}^{rad}=-1/3
    \eeq

The further study of such torsional Universes goes beyond the scope of this work and  the interested reader is referred  to \cite{kranas2019friedmann,Iosifidis:2021iuw} and also \cite{Barrow:2019bvx,Pereira:2019yhu} for more information on the topic. We should point out however that in these works the important relations (\ref{rho11}) and (\ref{p22}), along with the evolution equation (\ref{sigtor}) were missing and here these are revealed by the use of the proper Lagrangian formulation. This justifies the power and usefulness of the Action Principle.

\subsection{Pure Shear Cosmology}

Let us now examine the  pure Shear fluid in the Cosmological setting. Firstly, taking the totally antisymmetric part of the connection field equations (\ref{Cone}), it follows that $P=0$. Furthermore, we may choose the gauge (of the projective freedom) so as to obtain a vanishing torsion vector in this gauge, $S_{\mu}=0$ which implies $\Phi=0$ and as a result (see eq. (\ref{isotor})) the full torsion vanishes. Therefore, the pure shear part only sources non-metricity as it is expected. From the connection equations of motion, and for such a cosmological setting (recall the vanishing of torsion)  we finally find
\beq
B=-2\kappa \phi 
\eeq
\beq
A=-\kappa \phi
\eeq
\beq
C=-\kappa \phi -2 \kappa \psi
\eeq
and the conservation laws read, in this case,
\beq
p=\hat{p}+\frac{1}{2}\Big[ \dot{\phi}+H(3 \phi +2\psi) \Big] \label{pphath}
\eeq
\beq
\dot{\rho}+3 H(\rho+p)=-\kappa \phi (\rho-\hat{\rho}) \label{llk}
\eeq
Do also recall the relation
\beq
\rho-\hat{\rho}=3(p-\hat{p}) \label{rrrr}
\eeq
as a consequence of vanishing dilation. Note that $\Pi=\phi$. Now let us assume for simplicity that $\rho=\rho(n, \Pi)$ namely that the shear hyperfluid depends only on one of the shear variables (consequently $\psi=0$). In this instance we find
\beq
\frac{\partial \rho}{\partial \Pi}=-\frac{3 \kappa}{4}\Pi
\eeq
which implies that 
\beq
\rho=-\frac{3 \kappa}{8}\Pi^{2}+f(n)=-\frac{3 \kappa}{8}\phi^{2}+\rho_{pf}(n) \label{tgt}
\eeq
much like the pure spin fluid case. Correspondingly, the pressure is found to be
\beq
p=n\frac{\partial \rho}{\partial n}+\phi\frac{\partial \phi}{\partial \phi}-\rho=\Big( n \frac{\rho_{pf}}{d n}-\rho_{pf} \Big)-\frac{3 \kappa}{8}\phi^{2}=p_{pf}-\frac{3 \kappa}{8}\phi^{2} \label{tgt2}
\eeq

Furthermore, from the metric field equations one can then derive the modified Friedmann equations which we display here for completeness 
\beq
H^{2}=\frac{\kappa}{2}\Big[ \dot{\phi}+3 H \phi\Big]+\frac{\kappa^{2}}{4}\phi^{2}+\frac{\kappa \hat{\rho}}{3}
\eeq
\beq
\frac{\ddot{a}}{a}=-\frac{\kappa}{6}(\hat{\rho}+3 \hat{p})-\frac{\kappa^{2}}{2}\phi^{2}-\frac{\kappa}{2}\Big[ \dot{\phi}+3 H \phi \Big]
\eeq
With the help of (\ref{pphath})-(\ref{rrrr}) there are further simplified to (compare also with \cite{Iosifidis:2020upr})
\beq
H^{2}=\frac{\kappa \rho}{3}+\frac{\kappa^{2}}{4}\phi^{2} \label{gfd}
\eeq
\beq
\frac{\ddot{a}}{a}=-\frac{\kappa}{6}(\rho+3 p)-\frac{\kappa^{2}}{2}\phi^{2} \label{ikj}
\eeq
As a consistency check it can be easily confirmed that differentiating (\ref{gfd}) and using the conservation laws one arrives at (\ref{ikj}).
Let us note that the expressions for $\rho$ and $p$ above are not the purely perfect fluid contributions but have an inherent shear dependence as well, see equations (\ref{tgt}) and (\ref{tgt2}). Employing also the latter we exhaust all relations and arrive to the final form of the Friedmann equations 
\beq
H^{2}=\frac{\kappa \rho_{pf}}{3}+\frac{\kappa^{2}}{8}\phi^{2} \label{gfd}
\eeq
\beq
\frac{\ddot{a}}{a}=-\frac{\kappa}{6}(\rho_{pf}+3 p_{pf})-\frac{\kappa^{2}}{4}\phi^{2} \label{ikj}
\eeq
We have followed a didactic step by step derivation here in order to clarify all  subtle little points of the procedure. Differentiating the former of the above with respect to $t$ and using the latter we get
\beq
\dot{\rho}_{pf}+3 H(\rho_{pf}+p_{pf})=-\frac{3 \kappa}{4}\phi (\dot{\phi}+3 H \phi)
\eeq
which also follows immediately by direct substitution of (\ref{tgt}) and (\ref{tgt2}) into (\ref{llk}). From this point onward we can proceed in a way identical to the pure spin case, namely if we assume the ansatz $\phi=\mu_{1}H$ and combine equation (\ref{gfd}) with the last conservation law we get an equation for the Hubble which reads as follows
\beq
\dot{H}+\mu_{2}H^{2}=0
\eeq
where $\mu_{2}$ is a constant depending on $\mu_{1}$. One can consequently read-off the effective barotropic index $w_{eff}$ in this case as well, in a way quite similar to that of the pure spin case. Wrapping up, with these two examples we have demonstrated the consistency and applicability of our general construction of hyperhydrodynamics.

\section{Conclusions}
\label{conclu}

We have developed the action principle for relativistic hyperfluids. The latter constitute a generalization of relativistic fluids with the inclusion of hypermomentum which is associated to the micro-properties of matter (spin, dilation and shear). We found that the interaction terms that one needs to add to the fluid Lagrangian in order to generate nonvanishing hypermomentum tensor, generically introduce modifications of  the energy-momentum tensor making it acquire an imperfect fluid form . What is quite remarkable is that the modified energy-momentum tensor has the form that describes dissipative viscous fluids. Thus, on one side, the implications of hypermomentum have clear physical interpretation in hydrodynamics. On the other hand, the phenomenological parameterisations used to describe viscous effects in hydrodynamics could find a useful geometrical interpretation in terms of distortion of the affine connection,
and their more satisfactory theoretical derivation from an action principle. This approach, {\it hyperhydrodynamics}, suggests applications from condensed matters to cosmology. 

The so-called constitutive relations that are assumed in the development of dissipative fluids\footnote{In order to be compatible with the $2^{nd}$ law \cite{hiscock1983stability}.}, in our formulation are the consequence of the dynamical equations and need not be imposed by hand. Therefore, the kinematic coefficients (i.e. bulk and shear viscosity, thermal conductivity etc.) are directly related to the assumed micro-properties of the fluid as they follow from the inclusion of the hypermomentum. All the terms appearing in the constitutive relations of the hyperfluid model considered in section \ref{cosmo} have a well-known interpretation (with the appropriate correspondences) in the first-order development of relativistic viscous fluids \cite{bemfica2022first}. To reiterate, a particularly appealing characteristic of our development is that the coefficient functions are not just free functions but rather, they are all related to the hypermomentum degrees of freedom (microstructure), the relations being
determined by an underlying action principle. 

The form of the fluid action that may result in consistent hyperhydrodynamics, depends on the kinetic terms for the connection. In the context of gravity, this means that the gravitational theory determines the couplings that we may introduce to hypermomentum. Interestingly, the standard gravitational theory (i.e. general relativity) turns out to correspond to the expected (i.e. quadratic) form of the hyperfluid couplings and results in a standard description of imperfect fluids (i.e. the first-order development of viscous hydrodynamics). For instance, the action for spin hyperhydrodynamics
\begin{equation}
S = \frac{1}{2\kappa}\int d^n x \sqrt{-g}\left[R - f(n) + \frac{d-1}{4(d-2)}\kappa\sigma^2 + \frac{3}{4}\kappa\zeta^2 
+ \lp S_\mu - \frac{1}{4}q_\mu + \frac{1}{4}Q_\mu\rp \sigma u^\mu
+ \frac{1}{2}t_\mu \zeta u^\mu  
\right]
\end{equation}
leaves the freedom to adjust the perfect fluid sector of the fluid (the effect of number density $n$ on the energy density $\rho$), but the hypermomentous couplings (the effect of the spin densities $\sigma$ and $\zeta$) are dictated up to the numerical coefficients. The expressions (\ref{flux}) and (\ref{astress}) would seem to imply that the imperfect properties of the fluid induced by the spin part of hypermomentum become more important at low temperatures\footnote{By a first glance it seems that for a pure shear fluid the behaviour is reversed, namely the hypermomentum contribution increases with temperature. Of course in order to conclude something more well established and concrete one needs to have a specific model to work with.}, at least by a naive extrapolation of the standard interpretation of the coefficients in the fluid expansion. As it is clear from our derivations, summarised in section \ref{general}, for more generic gravity actions corresponding to modifications of Einstein gravity, the hyperfluid may assume quite non-standard forms which do not allow the direct comparison with the known parameterisations of imperfect fluids. By reverse-engineering, one may deduce the properties of hyperfluids which may be consistent with specific modifications (or reformulations) of gravity. In particular, it could be interesting to study hyperhydrodynamics in teleparallel equivalents of general relativity. Teleparallelism of affine geometry automatically results in the conservative property of the hyperfluid, which could facilitate the further development of the new formalism.  It would also be interesting to find how the entropy current \cite{Romatschke:2009kr,Loganayagam:2008is} is generalized in the case of hyperhydrodynamics and also find possible associations of the hyperfluid to the quark-gluon plasma \cite{BRAHMS:2004adc,Pasechnik:2016wkt,Muronga:2003ta}.



\section{Acknowledgements}

D. I.'s work was supported by the Estonian Research Council grant SJD14, and T.K.'s work was supported by the grant PRG356. This work was supported by the Center of Excellence TK202 "Foundations of the Universe".c 
D.I. would like to thank Anastasios Petkou for some very useful discussions.
 
\appendix

\section{Variations}

Let us gather here some useful variations. Firstly let us derive the variations of particle number density $n$. Recalling that
\beq
n:=\frac{|J|}{\sqrt{-g}} \label{nbasic}
\eeq
and the fact that $|J|:=\sqrt{-g_{\mu\nu}J^{\mu}J^{\nu}}$, we can easily find, for the g-variation of the latter
\beq
\delta_{g}n=(\delta g^{\mu\nu})\Big[\frac{n}{2}(g_{\mu\nu}+u_{\mu}u_{\nu}) \Big] \label{gn}
\eeq
In fact, similar formulas hold true for the g-variations of the rest of the hyperfluid densities. Let us generally denote a hypermomentum density by $\mathcal{X}$ where $\mathcal{X}=h,\sigma,\zeta,\Sigma_{1},\Sigma_{2}$ and the associated densitized flux vector 
\beq
\mathcal{X}^{\mu}=\sqrt{-g}\mathcal{X}u^{\mu}=\sqrt{-g}X^{\mu}
\eeq
then 
\beq
\mathcal{X}=\frac{|\mathcal{X}|}{\sqrt{-g}}
\eeq
and similarly to the above we find
\beq
\delta_{g}\mathcal{X}=(\delta g^{\mu\nu})\Big[\frac{\mathcal{X}}{2}(g_{\mu\nu}+u_{\mu}u_{\nu}) \Big] \label{gX}
\eeq
Therefore for any density, including the particle number density n, the above variation holds true. Let us collectively then label all 6 densities (5 from hypermomentum + 1 from particle number) by $\mathcal{X}_{I}=\{n,h,\sigma,\zeta,\Sigma_{1},\Sigma_{2}\}$, $I=1,2,...,6$ and denote the set of the associated current densities by $\mathcal{X}_{I}^{\mu}=\sqrt{-g}\mathcal{X}_{I}u^{\mu}$.

Then, it is worth noting that every densitized quantity is by construction metric-independent. To prove this statement, 
we then have to show that
\beq
\delta_{g}\mathcal{X}_{I}^{\mu}=\delta_{g}(\sqrt{-g}\mathcal{X}_{I}u^{\mu})=0
\eeq
for all $I=1,2,...,6$. By a direct computation and using (\ref{gX}) and the fact that $\delta_{g}\sqrt{-g}=-\frac{1}{2}\sqrt{-g}g_{\mu\nu}\delta g^{\mu\nu}$ we readily find
\beq
\delta_{g}(\sqrt{-g}\mathcal{X}_{I}u^{\mu})=\frac{1}{2}\sqrt{-g}\mathcal{X}_{I}(\delta g^{\mu\nu}) \Big[  u_{\mu}u_{\nu}u^{\alpha}+2\frac{\delta u^{\alpha}}{\delta g^{\mu\nu}} \Big] \label{ts}
\eeq
Therefore, for the variation to vanish, the term inside the parenthesis must be identically zero. This is indeed the case, since starting from the line element
\beq
d \tau^{2}=-g_{\alpha\beta} dx^{\alpha} dx^{\beta}
\eeq
it easily follows that (see also \cite{Haghani:2023uad})
\beq
2 \frac{\delta (d \tau)}{d \tau}=-(\delta g_{\mu\nu})u^{\mu}u^{\nu}
\eeq
such that
\beq
\delta u^{\alpha}=\delta \left( \frac{d x^{\alpha}}{d \tau} \right)=-u^{\alpha}\frac{\delta (d \tau)}{d \tau}=\frac{1}{2}u^{\alpha}u^{\mu}u^{\nu}(\delta g_{\mu\nu})
\eeq
and finally using the identity
\beq
\delta g_{\mu\nu}=-g_{\mu\alpha}g_{\nu\beta}(\delta g^{\alpha\beta})
\eeq
we obtain
\beq
\delta u^{\alpha}=-\frac{1}{2}u^{\alpha}u_{\mu}u_{\nu}\delta g^{\mu\nu} \label{dgu}
\eeq
which implies that the terms inside the parenthesis of (\ref{ts}) indeed cancel  out, leaving us with
\beq
\delta_{g}\mathcal{X}_{I}^{\mu}=\delta_{g}(\sqrt{-g}\mathcal{X}_{I}u^{\mu})=0
\eeq
as stated. So we have the relations
\beq
\delta_{g}J^{\mu}=0=\delta_{g}\Delta^{\mu}=\delta_{g}\sigma^{\mu}=\delta_{g}\zeta^{\mu}=\delta_{g}\Sigma^{\mu}=\delta_{g}\Pi^{\mu}
\eeq
It is also worth mentioning that ratios of densities with particle number are also metric independent. So under a redefinition 
\beq
\mathcal{X} \rightarrow \mathcal{\bar{X}}=\frac{\mathcal{X}}{n}
\eeq
of the density $\mathcal{X}$, the new density $\mathcal{\bar{X}}$ has zero metric variation, i.e.
\beq
\delta_{g}\mathcal{\bar{X}}=0
\eeq
as can be easily seen by using the above results for the densities variations. This actual means that if we start we the density as the function $\rho(n,\mathcal{\bar{X}})$ and the associated 'hyperpressure'
\beq
p=n\frac{\partial \rho}{\partial n}+\mathcal{\bar{X}}\frac{\partial \rho}{\partial \mathcal{\bar{X}}}-\rho
\eeq
then under the rescaling $\mathcal{\bar{X}}$ we have that $\rho(n,\mathcal{\bar{X}}) \rightarrow \rho(n,\frac{\mathcal{X}}{n})$ and by using the chain rule it follows that
\beq
n\frac{\partial \rho}{\partial n}\rightarrow n\frac{\partial \rho}{\partial n}-\mathcal{\bar{X}}\frac{\partial \rho}{\partial \mathcal{\bar{X}}}
\eeq
and consequently the above hyperpressure transforms to
\beq
p\rightarrow p=n\frac{\partial \rho}{\partial n}-\rho \label{ppp}
\eeq
namely in the new rescaled variables the definition for the pressure becomes the standard one. As a result one can either start with $\rho(n,\mathcal{X}/n)$ right from the on-set and perform the variations or alternatively start with $\rho(n, \mathcal{X})$ perform the variations and substitute $\mathcal{X}\rightarrow n\mathcal{X}$ in the end along with the replacement (\ref{ppp}).

Using (\ref{dgu})) we can easily find also the variation of $u_{\lambda}=g_{\lambda\nu}u^{\nu}$ which reads
\beq
\frac{\delta u_{\lambda}}{\delta g^{\mu\nu}}=-\frac{1}{2}(u_{\mu}g_{\nu\lambda}+u_{\nu}g_{\mu\lambda}+u_{\lambda}u_{\mu}u_{\nu})
\eeq
where the identity 
\beq
\frac{\delta g^{\alpha\beta}}{\delta g^{\mu\nu}}=\delta^{(\alpha}_{\mu}\delta^{\beta)}_{\nu}
\eeq
has been employed.
Also, useful relations can be obtained
if we vary $u_{\mu}u_{\nu}g^{\mu\nu}=u_{\mu}u^{\mu}=-1$. Doing so  we get the relations
\beq
2 u^{\alpha} \frac{\delta u_{\alpha}}{\delta g^{\mu\nu}}=-u_{\mu} u_{\nu}
\eeq
\beq
 u^{\alpha} \frac{\delta u_{\alpha}}{\delta g^{\mu\nu}}=- u_{\alpha} \frac{\delta u^{\alpha}}{\delta g^{\mu\nu}}
\eeq
\beq
2 u_{\alpha} \frac{\delta u^{\alpha}}{\delta g^{\mu\nu}}=u_{\mu} u_{\nu}
\eeq
which of course also follow by simply contracting the above derived variations of $u^{\mu}$ and $u_{\lambda}$. Using relations (\ref{gn}) and (\ref{gX}) we easily find the variation of the density $\rho(n,h,\sigma,\zeta,...)=\rho(\mathcal{X}_{I})$ with respect to the metric,
\beq
\frac{\delta \rho}{\delta g^{\mu\nu}}=\frac{1}{2}(g_{\mu\nu}+u_{\mu}u_{\nu})\sum_{I} \mathcal{X}_{I} \frac{\partial \rho}{\partial \mathcal{X}_{I}} 
\eeq

Next, let us compute the variation
\beq
\frac{\delta \rho}{\delta J^{\mu}}
\eeq
Note that $\rho$ implicitly depends on $J^{\mu}$ due to  relation (\ref{nbasic}). Then, employing the chain rule we have 
\beq
\frac{\delta \rho}{\delta J^{\mu}}=\frac{\partial \rho}{\partial n}\frac{\delta n}{\delta J^{\mu}}
\eeq
and by varying (\ref{nbasic}) we respect to $J^{\mu}$ we trivially find
\beq
\frac{\delta n}{\delta J^{\mu}}=-\frac{J_{\mu}}{|J|\sqrt{-g}}=-\frac{u_{\mu}}{\sqrt{-g}}
\eeq
and as a result
\beq
\frac{\delta \rho}{\delta J^{\mu}}=-\frac{\partial \rho}{\partial n} \frac{u_{\mu}}{\sqrt{-g}}=-\frac{1}{\sqrt{-g}}\mu u_{\mu}
\eeq
Of course the latter holds true not only for $J^{\mu}$ but also for all of the hypermomentum current densities, and an identical calculation reveals that
\beq
\frac{\delta \rho}{\delta \mathcal{X}^{\mu}}=-\frac{1}{\sqrt{-g}}\frac{\partial \rho}{\partial \mathcal{X}}u_{\mu}
\eeq
Using the above variations it is crucial to observe that
\beq
\frac{\delta}{\delta J^{\mu}}\left( \frac{J^{\alpha}}{n}\right)=\frac{1}{n}(\delta^{\alpha}_{\mu}+u_{\mu}u^{\alpha})=\frac{1}{n}h^{\alpha}{}_{\mu}
\eeq
In particular it implies that the $J^{\mu}$-variation of $ J^{\alpha}/n$ is orthogonal to both $J^{\mu}$ and $J_{\alpha}$. In addition it holds that
\beq
\frac{|J|}{\delta J^{\mu}}=-\frac{J_{\mu}}{|J|}=-u_{\mu}
\eeq
as well as
\beq
\frac{\delta u^{\alpha}}{\delta J^{\mu}}=\frac{1}{|J|}h^{\alpha}{}_{\mu}
\eeq

For the metric variations of the connection couplings we have the following formulae (up to total derivatives)
\beq
\delta_{g}(Q_{\mu}\mathcal{D}^{\mu})=-(\delta g^{\mu\nu})g_{\mu\nu}\partial_{\alpha}\mathcal{D}^{\alpha}
\eeq

\beq
\delta_{g}(t_{\mu}\zeta^{\mu})=\zeta^{\mu}\delta_{g}t_{\mu}=(\delta g^{\mu\nu})\Big[  -\frac{1}{2}(\zeta^{\alpha}t_{\alpha})g_{\mu\nu}+ 2 \zeta^{\rho}\epsilon_{\rho\alpha\beta(\mu}S_{\nu)}{}^{\alpha\beta} \Big]
\eeq
\beq
\delta_{g}(\sigma^{\nu}q_{\nu})=\sigma^{\nu}\delta_{g}q_{\nu}=(\delta g^{\mu\nu})\Big[ -q_{(\mu}\sigma_{\nu)}+2 S_{(\mu}\sigma_{\nu)}-\nabla_{(\mu}\sigma_{\nu)} \Big]+t.d.
\eeq

\section{Useful Relations}
Considering the usual $1+(d-1)$ spacetime split we have the timelike d-velocity field as the tangent vector to a curve, which reads
\beq
u^{\mu}:=\frac{d x^{\mu}}{d \lambda}
\eeq
In general, for arbitrary parameter $\lambda$  the latter is not  normalized to $- 1$ (see \cite{Iosifidis:2018diy}  ). However, if $\lambda$ is taken to be the proper time $\tau$ then the velocity is indeed normalized to $-1$ by default, namely
\beq
u_{\mu}u^{\mu}=-1
\eeq
when 
\beq
u^{\mu}:=\frac{d x^{\mu}}{d \tau}
\eeq
We shall therefore consider the affine parameter to be the proper time throughout. Then we may define the usual projector
\beq
h_{\mu\nu}=g_{\mu\nu}+ u_{\mu} u_{\nu}
\eeq
We may then define the hyper-acceleration as
\beq
a_{\mu}:=u^{\alpha}\nabla_{\alpha}u_{\mu}
\eeq
and performing a post-Riemannian expansion it is split into
\beq
a_{\mu}=\tilde{a}_{\mu}-N_{\alpha\mu\beta}u^{\alpha}u^{\beta} \label{accrel}
\eeq
or in terms of torsion and non-metricity
\beq
a_{\mu}=\tilde{a}_{\mu}-\frac{1}{2}Q_{\mu\alpha\beta}u^{\alpha}u^{\beta}  -2 S_{\mu\alpha\beta}u^{\alpha}u^{\beta}  \label{accrel2}
\eeq

where $\tilde{a}_{\lambda}$ is the usual Riemannian acceleration. Let us note that the above generalized acceleration is not orthogonal to the velocity. Indeed, acting on $u_{\mu}u^{\mu}=-1$ one time with $\tilde{\nabla}$ and another with $\nabla$ we get the identities
\beq
u^{\mu}\tilde{\nabla}_{\lambda}u_{\mu}=0
\eeq
and
\beq
2 u^{\mu}\nabla_{\alpha}u_{\mu}=-Q_{\alpha\mu\nu}u^{\mu}u^{\nu}
\eeq
respectively. Contracting the former with $u^{\lambda}$ immediately follows that $u_{\mu}\tilde{a}^{\mu}=0$ indicating that the Riemannian acceleration is orthogonal to the velocity, while if we contract the latter with $u^{\lambda}$ we see that
\beq
a_{\mu}u^{\mu}=u^{\mu}u^{\nu}\nabla_{\mu}u_{\nu}=-\frac{1}{2}Q_{\alpha\mu\nu}u^{\alpha}u^{\mu}u^{\nu}
\eeq
namely they are not orthogonal, as stated.

Now, the usual notions of expansion, shear and rotation of a fluid can be generalized also to non-Riemannian backgrounds \cite{Iosifidis:2018diy}. For instance one may define the non-Riemannian expansion scalar as
\beq
\Theta:=g^{\mu\nu}\nabla_{\mu}u_{\nu}
\eeq
However, in order to be able to compare with the usual Riemannian case it is beneficial to express all quantities into usual Riemannian and post-Riemannian contributions of torsion and non-metricity. Doing so for the above scalar we trivially find
\beq
\Theta=\tilde{\Theta}+A_{\mu}u^{\mu}
\eeq
where $\tilde{\Theta}:=\tilde{\nabla}_{\mu}u^{\mu}$ is the usual Riemannian expansion scalar and
\beq
A_{\mu}=2 S_{\mu}+\frac{Q_{\mu}}{2}-q_{\mu}
\eeq
More generally we may expand
\beq
\nabla_{\mu}u_{\nu}=\tilde{\nabla}_{\mu}u_{\nu}-N^{\lambda}{}_{\nu\mu}u_{\lambda} \label{nt}
\eeq
and use the well-known split
\beq
\tilde{\nabla}_{\beta}u_{\alpha}=\tilde{\sigma}_{\alpha\beta}+\frac{\tilde{\Theta}}{(d-1)}h_{\alpha\beta}+\tilde{\omega}_{\alpha\beta}-\tilde{a}_{\alpha}u_{\beta}
\eeq
with
\beq
\tilde{\Theta}:=h^{\alpha\beta}\tilde{\nabla}_{\alpha}u_{\beta}=g^{\alpha\beta}\tilde{\nabla}_{\alpha}u_{\beta}
\eeq
\beq
\tilde{\sigma}_{\mu\nu}:=h^{\alpha}_{(\mu}h^{\beta}_{\nu)}\tilde{\nabla}_{\beta}u_{\alpha}-\frac{\tilde{\Theta}}{(d-1)}h_{\mu\nu}
\eeq
\beq
\tilde{\omega}_{\mu\nu}:=h^{\alpha}_{[\mu}h^{\beta}_{\nu]}\tilde{\nabla}_{\beta}u_{\alpha}
\eeq
representing the usual Riemannian, expansion, shear and vorticity of the fluid respectively, to write
\beq
\nb_{[\mu} u_{\nu]}=-\tilde{\omega}_{\mu\nu}-u_{[\mu}\tilde{a}_{\nu]}+S_{\mu\nu\alpha}u^{\alpha} \label{vort}
\eeq
and
\beq
\nb_{(\mu} u_{\nu)}=\tilde{\sigma}_{\mu\nu}+\frac{\tilde{\Theta}}{(d-1)}h_{\mu\nu}-\tilde{a}_{(\mu}u_{\nu)}-N_{\alpha(\mu\nu)}u^{\alpha}
\eeq

Let us also include here  some metric variations that are useful in deriving the form of the energy-momentum tensor. For an arbitrary scalar function $f$ and the associated  vector density $f^{\alpha}:=\sqrt{-g}f u^{\alpha}$, it holds that
\beq
\delta_{g}(Q_{\alpha\mu\nu}f^{\alpha}u^{\mu}u^{\nu})=\delta g^{\mu\nu} \Big[ -(\partial_{\alpha} f^{\alpha}+f^{\alpha}Q_{\alpha\beta\gamma}u^{\beta}u^{\gamma})u_{\mu}u_{\nu}-2 f_{(\mu}a_{\nu)}-2 f^{\alpha}Q_{\alpha\beta(\mu}u_{\nu)}u^{\beta}  \Big]
\eeq
\beq
\delta_{g}(f^{\alpha}q_{\alpha})=f^{\alpha}\delta_{g}q_{\alpha}=\delta g^{\mu\nu}\Big[ A_{(\mu}f_{\nu)}-\sqrt{-g} u_{(\mu}\partial_{\nu)}f -\sqg f \nabla_{(\mu}u_{\nu)} \Big]
\eeq

The following relations are useful in deriving the form of the canonical energy-momentum tensor:
\beq
g^{\lambda\nu}(2 S_{\alpha}-\nabla_{\alpha})(\sqrt{-g}f \delta_{\lambda}^{\mu}f^{\alpha})=-g^{\mu\nu}(\partial_{\alpha}f^{\alpha})
\eeq
\beq
g^{\lambda\nu}(2 S_{\alpha}-\nabla_{\alpha})(\sqrt{-g}f u_{\lambda}u^{\mu}u^{\alpha})=-(\partial_{\alpha}f^{\alpha})u^{\mu}u^{\nu}+\sqrt{-g}f\Big[  -2 \tilde{a}^{(\mu}u^{\nu)}+N^{\alpha\nu\beta}u_{\alpha}u_{\beta}u^{\mu}-N^{\mu}{}_{\alpha\beta}u^{\alpha}u^{\beta}u^{\nu}\Big]
\eeq
\beq
g^{\lambda\nu}(2 S_{\alpha}-\nabla_{\alpha})(\sqrt{-g}f u^{\mu}\delta^{\alpha}_{\lambda})=-\sqg u^{\mu}\partial^{\nu}f +f \sqg \Big[  A^{\nu}u^{\mu}+q^{\nu}u^{\mu}-\tilde{\nb}^{\nu}u^{\mu}+N^{\mu\alpha\nu}u_{\alpha} \Big]
\eeq
\beq
g^{\lambda\nu}(2 S_{\alpha}-\nabla_{\alpha})(\sqrt{-g}f u_{\lambda}g^{\mu\alpha})=-\sqg u^{\nu}\partial^{\mu}f +f \sqg \Big[  A^{\mu}u^{\nu}-\tilde{\nb}^{\mu}u^{\nu}+N^{\alpha\nu\mu}u_{\alpha} \Big]
\eeq

\beq
g^{\lambda\nu}(2 S_{\alpha}-\nabla_{\alpha})(\sqrt{-g}f (u^{\mu}\delta^{\alpha}_{\lambda}+u_{\lambda}g^{\mu\alpha})=\sqrt{-g}\Big( 2S^{(\mu}+\frac{Q^{(\mu}}{2}\Big) 2f u^{\nu)}-\sqrt{-g}2 \partial^{(\mu}f u^{\nu)}-\sqrt{-g}f \Big[ g^{\lambda\nu}\nabla_{\lambda}u^{\mu}+g^{\lambda\nu}g^{\mu\alpha}\nabla_{\alpha}u_{\lambda}+u^{\nu}q^{\mu}\Big]
\eeq
or expanding the covariant derivatives
\beq
g^{\lambda\nu}(2 S_{\alpha}-\nabla_{\alpha})(\sqrt{-g}f (u^{\mu}\delta^{\alpha}_{\lambda}+u_{\lambda}g^{\mu\alpha})=-\sqrt{-g}2 u^{(\mu}\partial^{\nu)}f +\sqrt{-g}f \Big[ 2 A^{(\mu}u^{\nu)} -2 \tilde{\nabla}^{(\mu}u^{\nu)}+u^{\mu}q^{\nu}+(N^{\alpha\nu\mu}-N^{\mu\alpha\nu})u_{\alpha}\Big]
\eeq
and
\beq
g^{\lambda\nu}(2 S_{\alpha}-\nabla_{\alpha})(\sqrt{-g}f (u^{\mu}\delta^{\alpha}_{\lambda}-u_{\lambda}g^{\mu\alpha})=\sqg 2 u^{[\nu}\partial^{\mu]}f+ \sqg f \Big[ 2 u^{[\mu}A^{\nu]}+2  \tilde{\nabla}^{[\mu}u^{\nu]}+q^{\nu}u^{\mu}-(N^{\alpha\nu\mu}+N^{\mu\alpha\nu})u_{\alpha} \Big]
\eeq

We also compute
\beq
(2 S_{\nu}-\nabla_{\nu})(\sqrt{-g}\epsilon_{\lambda}{}^{\mu\nu\rho}u_{\rho}\zeta)=\sqrt{-g}\epsilon_{\lambda}{}^{\mu\alpha\beta}\Big[ 2S_{[\alpha}u_{\beta]}\zeta-\zeta \nabla_{[\alpha}u_{\beta]}- u_{[\beta}\partial_{\alpha]} \zeta\Big]+\sqrt{-g}\epsilon^{\mu\alpha\beta\gamma}Q_{\alpha\beta\lambda}u_{\gamma}\zeta
\eeq
or by employing (\ref{nt})
\beq
-\frac{1}{\sqrt{-g}}(2 S_{\nu}-\nabla_{\nu})(\sqrt{-g}\epsilon_{\lambda}{}^{\mu\nu\rho}u_{\rho}\zeta)=-\epsilon_{\lambda}{}^{\mu\alpha\gamma}2 S_{\alpha}u_{\gamma}\zeta +S_{\alpha\beta}{}{}^{\gamma}u_{\gamma}\epsilon_{\lambda}{}^{\mu\alpha\beta}\zeta +\epsilon_{\lambda}{}^{\mu\alpha\beta} \Big(\zeta \tilde{\nb}_{[\alpha}u_{\beta]}+u_{[\beta}\partial_{\alpha]}\zeta \Big) -\epsilon^{\mu\alpha\beta\gamma}Q_{\alpha\beta\lambda}u_{\gamma}\zeta
\eeq
Note that due to the identity
\beq
-\frac{1}{\sqrt{-g}}(2 S_{\nu}-\nabla_{\nu})(\sqrt{-g}\epsilon_{\lambda}{}^{\mu\nu\rho}u_{\rho}\zeta)=\tilde{\nb}_{\alpha}(\epsilon_{\lambda}{}^{\mu\alpha\gamma}u_{\gamma}\zeta)+N^{\mu}{}_{\alpha\beta}\epsilon_{\lambda}{}^{\alpha\beta\gamma}u_{\gamma}\zeta -N^{\alpha}{}_{\lambda\beta}\epsilon_{\alpha}{}^{\mu\beta\gamma}u_{\gamma}\zeta
\eeq
the above expression should agree with its post-Riemannian expanded version
\begin{gather}
    \tilde{\nb}_{\alpha}(\epsilon_{\lambda}{}^{\mu\alpha\gamma}u_{\gamma}\zeta)+N^{\mu}{}_{\alpha\beta}\epsilon_{\lambda}{}^{\alpha\beta\gamma}u_{\gamma}\zeta -N^{\alpha}{}_{\lambda\beta}\epsilon_{\alpha}{}^{\mu\beta\gamma}u_{\gamma}\zeta= \nonumber \\
    =\epsilon_{\lambda}{}^{\mu\alpha\beta} \Big(\zeta \tilde{\nb}_{[\alpha}u_{\beta]}+u_{[\beta}\partial_{\alpha]}\zeta \Big) -\epsilon^{\mu\alpha\beta\gamma}Q_{\alpha\beta\lambda}u_{\gamma}\zeta +S_{\alpha\beta}{}{}^{\mu}\epsilon_{\lambda}{}^{\alpha\beta\gamma}u_{\gamma}\zeta +S_{\alpha\beta\lambda}\epsilon^{\alpha\mu\beta\gamma}u_{\gamma}\zeta
\end{gather}
and this happens iff
\beq
-\epsilon_{\lambda}{}^{\mu\alpha\gamma}2 S_{\alpha}u_{\gamma} +S_{\alpha\beta}{}{}^{\gamma}u_{\gamma}\epsilon_{\lambda}{}^{\mu\alpha\beta}=S_{\alpha\beta}{}{}^{\mu}\epsilon_{\lambda}{}^{\alpha\beta\gamma}u_{\gamma}+S_{\alpha\beta\lambda}\epsilon^{\alpha\mu\beta\gamma}u_{\gamma} \label{ident}
\eeq
Therefore, to check that are calculations are indeed correct we must show that the latter holds as an identity and not as a constraint. Let us show that this is indeed the case, namely that (\ref{ident}) is identically satisfied. To do so we start by the dimensional-dependent identity
\beq
g^{\mu\nu}\epsilon_{\lambda}{}^{\alpha\beta\gamma}=\delta_{\lambda}^{\nu}\epsilon^{\mu\alpha\beta\gamma}+g^{\nu\gamma}\epsilon_{\lambda}{}^{\mu\alpha\beta}+g^{\nu\alpha}\epsilon_{\lambda}{}^{\mu\beta\gamma}+g^{\nu\beta}\epsilon_{\lambda}{}^{\mu\gamma\alpha}
\eeq
Contracting with $S_{\alpha\beta\nu}$ we get
\beq
S_{\alpha\beta}{}{}^{\mu}\epsilon_{\lambda}{}^{\alpha\beta\gamma}+S_{\alpha\beta\lambda}\epsilon^{\alpha\mu\beta\gamma}=S_{\alpha\beta}{}{}^{\gamma}\epsilon_{\lambda}{}^{\mu\alpha\beta}-2 S_{\alpha} \epsilon_{\lambda}{}^{\mu\alpha\gamma}
\eeq
and a further contraction with $u_{\gamma}$ establishes (\ref{ident}) guaranteeing therefore that our computations are indeed correct. 

After using the expression for the vorticity we arrive at the final form
\begin{gather}
    -g^{\lambda\nu}\frac{1}{\sqrt{-g}}(2 S_{\alpha}-\nabla_{\alpha})(\sqrt{-g}\epsilon_{\lambda}{}^{\mu\nu\rho}u_{\rho}\zeta)=\epsilon^{\mu\nu\alpha\beta}\Big[ 
\zeta (2 S_{[\alpha}-\tilde{a}_{[\alpha})u_{\beta]}+u_{[\alpha}\partial_{\beta]}\zeta -S_{\alpha\beta}{}{}^{\gamma}u_{\gamma}\zeta\Big] +\zeta \epsilon^{\mu\nu\alpha\beta}\tilde{\omega}_{\alpha\beta}-\epsilon^{\mu\alpha\beta\gamma}Q_{\alpha\beta}{}{}^{\nu}u_{\gamma}\zeta
\end{gather}

\bibliography{hyperfluidrefs}

\end{document}